\pdfoutput=1
\documentclass[fp,12pt,twocolumn]{jpsj3}
\usepackage[dvipdfmx]{graphicx}
\usepackage{dcolumn}
\usepackage{bm}
\usepackage{color}
\usepackage{amsmath,amssymb}
\usepackage{braket}
\usepackage{booktabs}
\begin{document}
\setlength{\parskip}{0pt}
\setlength{\baselineskip}{13.5pt}

\newcommand{\non}{\nonumber}
\newcommand{\Tc}{T_{\mathrm{c}}}
\newcommand{\Tcopt}{T_\mathrm{c}^{\mathrm{opt}}}
\newcommand{\Tcmax}{T_\mathrm{c}^{\mathrm{max}}}
\newcommand{\Ns}{N_{\mathrm{s}}}
\newcommand{\bR}{\mbox{\boldmath $R$}}
\newcommand{\txr}[1]{\textcolor{black}{#1}}
\newcommand{\txrs}[1]{\textcolor{black}{\sout{#1}}}
\newcommand{\txm}[1]{\textcolor{black}{#1}}
\newcommand{\trs}[1]{\textcolor{black}{\sout{#1}}}
\newcommand{\tb}[1]{\textcolor{black}{#1}}
\newcommand{\tby}[1]{\textcolor{black}{#1}}
\newcommand{\tbs}[1]{\textcolor{black}{\sout{#1}}}
\newcommand{\Ha}{\mathcal{H}}
\newcommand{\mh}{\mathsf{h}}
\newcommand{\mA}{\mathsf{A}}
\newcommand{\mB}{\mathsf{B}}
\newcommand{\mC}{\mathsf{C}}
\newcommand{\mS}{\mathsf{S}}
\newcommand{\mU}{\mathsf{U}}
\newcommand{\mX}{\mathsf{X}}
\newcommand{\sP}{\mathcal{P}}
\newcommand{\sL}{\mathcal{L}}
\newcommand{\sO}{\mathcal{O}}
\newcommand{\la}{\langle}
\newcommand{\ra}{\rangle}
\newcommand{\ga}{\alpha}
\newcommand{\gb}{\beta}
\newcommand{\gc}{\gamma}
\newcommand{\gs}{\sigma}
\newcommand{\vk}{{\bm{k}}}
\newcommand{\vq}{{\bm{q}}}
\newcommand{\vR}{{\bm{R}}}
\newcommand{\vQ}{{\bm{Q}}}
\newcommand{\vga}{{\bm{\alpha}}}
\newcommand{\vgc}{{\bm{\gamma}}}
\newcommand{\avrg}[1]{\left\langle #1 \right\rangle}
\newcommand{\eqsa}[1]{\begin{eqnarray} #1 \end{eqnarray}}
\newcommand{\eqwd}[1]{\begin{widetext}\begin{eqnarray} #1 \end{eqnarray}\end{widetext}}
\newcommand{\hatd}[2]{\hat{ #1 }^{\dagger}_{ #2 }}
\newcommand{\hatn}[2]{\hat{ #1 }^{\ }_{ #2 }}
\newcommand{\wdtd}[2]{\widetilde{ #1 }^{\dagger}_{ #2 }}
\newcommand{\wdtn}[2]{\widetilde{ #1 }^{\ }_{ #2 }}
\newcommand{\cond}[1]{\overline{ #1 }_{0}}
\newcommand{\conp}[2]{\overline{ #1 }_{0#2}}
\newcommand{\nn}{\nonumber\\}
\newcommand{\cdt}{$\cdot$}
\newcommand{\bvec}[1]{\mbox{\boldmath$#1$}}
\newcommand{\blue}[1]{{#1}}
\newcommand{\bl}[1]{{#1}}
\newcommand{\red}[1]{\textcolor{black}{#1}}
\newcommand{\rr}[1]{{#1}}
\newcommand{\bu}[1]{\textcolor{black}{#1}}
\newcommand{\cyan}[1]{\textcolor{black}{#1}}
\newcommand{\fj}[1]{{#1}}
\newcommand{\green}[1]{{#1}}
\newcommand{\tm}[1]{\textcolor{black}{#1}}
\newcommand{\tgr}[1]{\textcolor{black}{#1}}
\newcommand{\tgrs}[1]{\textcolor{black}{\sout{#1}}}
\newcommand{\tmg}[1]{\textcolor{black}{#1}}
\newcommand{\tmgs}[1]{\textcolor{black}{\sout{#1}}}
\newcommand{\txc}[1]{\textcolor{black}{#1}}

\newcommand{\gr}[1]{\textcolor{black}{#1}}

\definecolor{green}{rgb}{0,0.5,0.1}
\definecolor{green1}{rgb}{0,1.0,0.0}
\definecolor{blue}{rgb}{0,0,0.8}
\definecolor{cyan}{rgb}{0,0.8,0.9}
\definecolor{grey}{rgb}{0.3,0.3,0.3}
\definecolor{orange}{rgb}{1,0.5,0.25}
\definecolor{purple}{rgb}{0.45,0.0,0.5}
\newcommand{\txo}[1]{\textcolor{black}{#1}}
\newcommand{\txp}[1]{\textcolor{black}{#1}}
\newcommand{\txcy}[1]{\textcolor{black}{#1}}

\title{
Superconductivity in bilayer $t$-$t'$ Hubbard models
}
\author{Akito Iwano$^{1,}$\thanks{iwano-akito975@g.ecc.u-tokyo.ac.jp} and Youhei Yamaji$^{2,}$\thanks{YAMAJI.Youhei@nims.go.jp}}
\inst{$^1$Department of Applied Physics, The University of Tokyo, Hongo, Bunkyo-ku, Tokyo, 113-8656, Japan \\
$^2$Center for Green Research on Energy and Environmental Materials, National Institute for Materials Science, Namiki, Tsukuba-shi, Ibaraki, 305-0044, Japan \\
} 
\date{\today}

\abst{
It has been a challenge in condensed matter physics to find superconductors with higher critical temperatures. 
Relationship between crystal structures and superconducting critical temperatures has attracted
considerable attention as a clue to designing higher-critical-temperature superconductors. In particular, the
relationship between the number $n$ of CuO$_2$ layers in a unit cell of copper oxide superconductors and the
optimum superconducting transition temperature $T_{\rm c}^{\rm opt}$ is intriguing. As experimentally observed in Bi, Tl,
and Hg based layered cuprates, $T_{\rm c}^{\rm opt}$ increases when the number of CuO$_2$ layers in the unit cell, $n$, is increased, 
up to $n = 3$, and, then, decreases for larger $n$. However, the mechanism behind the $n$ dependence of $T_{\rm c}^{\rm opt}$ remains 
elusive although there have been experimental and theoretical studies on the $n$ dependence. In
this paper, we focused on one of the simplest effective hamiltonians of the multilayer cuprates to clarify the
effects of the adjacent CuO$_2$ layers on the stability of the superconductivity. By utilizing a highly flexible
many-variable variational Monte Carlo method, we studied a bilayer $t$-$t'$ Hubbard model, in comparison
with the single layer $t$-$t'$ Hubbard model. Because the direct and quantitative simulation of $T_{\rm c}^{\rm opt}$ is still 
beyond the reach of the existing numerical algorithms, observables that correlate with $T_{\rm c}^{\rm opt}$ are examined 
in the present paper. Among the observables correlated with $T_{\rm c}^{\rm opt}$, the superconducting correlation at long 
distance and zero temperature is one of the easiest to calculate in the variational Monte Carlo method.
The amplitude of the superconducting gap functions is also estimated from the momentum distribution.
It is found that the in-plane superconducting correlation is not enhanced in comparison with the superconducting 
correlation in the single-layer $t$-$t'$ Hubbard model. While the superconducting correlations at
long distance both in the single-layer and bilayer models are almost the same at the optimal doping, the
superconducting correlations of the bilayer hamiltonian are significantly small in the overdoped region in
comparison with the correlations of the single-layer hamiltonian. The reduction at the overdoped region is
attributed to the van Hove singularity. In addition, we found that the amplitude of the superconducting
gap functions is also similar in both the single-layer and bilayer $t$-$t'$ Hubbard model at the optimal doping.
Therefore, we conclude that the adjacent Hubbard layers are not relevant to the enhancement of $T_{\rm c}^{\rm opt}$ in
the bilayer cuprates. Possible origins other than the adjacent layers are also discussed.
} 
\maketitle
\thispagestyle{plain}

\section{Introduction} \label{sec:SCinStrgsys} 
In condensed matter physics,
high-$T_\mathrm{c}$ superconductivity  that occurs in strongly correlated electron systems is
one of the central issues.
The mechanism of high-$T_\mathrm{c}$ superconductivity and
key factors that determine 
transition temperature ($\Tc$) have been puzzles to be solved.
\txc{The} high-$\Tc$ superconductors
generally mean materials that 
show higher $\Tc$ than that of conventional \txc{Bardeen-Cooper-Schrieffer} (BCS) superconductors,
at most around \txc{40 K},
or materials that show higher $\Tc$ than the liquid-nitrogen temperature ($\sim$ 77 K).
Regarding to the latter case,
the higher $\Tc$ than the liquid-nitrogen temperature at ambient pressure has been found only for copper oxide (cuprate) superconductors.
The cuprate High-$\Tc$ superconductors was first discovered by Bednorz and Müller in 1986 \cite{bednorz1986possible}. 
Although the $\Tc$ of the first cuprate superconductor is only about 30 K,
this discovery triggered a large number of studies to search for new materials
and led to discovery of various type of materials that show higher $\Tc$.
The maximum $\Tc$ found in the cuprates is 135 K at ambient pressure~\cite{schilling1993superconductivity},
which increases up to $\sim$ 160 K under high pressure~\cite{gao1994superconductivity}.
This maximum $\Tc$ is also the highest record among superconductors in transition metal compounds or other strongly correlated materials.

\tby{Cuprates superconductors share
the layered perovskite structure
and show anisotropic superconductivity
when electrons or holes are doped into the two-dimensional CuO$_2$ layers.
The anisotropic superconducting gap $\Delta(\bm{k})$ has $d_{x^2-y^2}$-wave symmetry\txo{~\cite{sigrist1992paramagnetic,van1995phase}},
\txc{which is often modeled by} $\Delta(\bm{k})\propto (\cos{k_x}-\cos{k_y})$.}

\tby{In contrast to these common features of cuprates,
$T_{\rm c}$ significantly depends on detailed crystal structures.}
\txc{Apical oxygen heights from CuO$_2$ planes have been known to correlate with the critical temperatures~\cite{PhysRevB.43.2968}.}
The correlation between
the number of the $\mathrm{CuO_2}$ layers and $\Tc^{\rm opt}$
in the
\txo{Bi~\cite{Maeda_1988},
Tl~\cite{sheng1988superconductivity}, and
Hg~\cite{putilin1991new} based}
homologous series of the hole-doped multilayer cuprates,
Bi$_{2}$Sr$_{2}$Ca$_{n-1}$Cu$_{n}$O$_{2n+4+\delta}$ [Bi$22(n\mathchar`-1)n$],
Tl$_{2}$Ba$_{2}$Ca$_{n-1}$Cu$_{n}$O$_{2n+4+\delta}$ [Tl$22(n\mathchar`-1)n$],
and
HgBa$_{2}$Ca$_{n-1}$Cu$_{n}$O$_{2n+2+\delta}$ [Hg$12(n\mathchar`-1)n$]~\cite{uchida2014high}.
In particular, the trilayer Hg-based cuprate Hg-1223 has the highest $\Tc$ mentioned above~\cite{SCOTT1994239,iyo2006synthesis}.
As explained in detail in the following section,
it has been universally known that $\Tc$ increases by increasing the number of the $\mathrm{CuO_2}$ layers in the unit cell, $n$, up to $n=3$.
Once $\Tc$ increases for $n\leq 3$ and shows maximum at $n=3$, and decreases monotonically for $n\geq4$. 


Although there are several theoretical studies~\cite{chakravarty1993interlayer,chakravarty1998electrons,leggett1999cuprate,chakravarty2004explanation,nishiguchi2013superconductivity,zegrodnik2017effect,medhi2007coexistence,okamoto2008enhanced}
to explain the layer number dependence of \txo{$\Tc^{\rm opt}$},
no scenarios have succeeded \tby{to quantitatively clarify} the dependence so far
and microscopic understanding is highly desirable \tby{to design cuprate
superconductors with higher critical temperatures}.
In this study, we concentrate on the simplest bilayer system and perform
numerical simulation of superconducting correlations
in the bilayer Hubbard model with single-particle hoppings between two adjacent layers.

\txo{We studied a bilayer $t$-$t'$ Hubbard model (see Sec.~\ref{Sec:Model_and_Methods}),
in comparison with the single layer $t$-$t'$ Hubbard model
by using a many-variable variational Monte Carlo method (reviewed in Sec.~\ref{Sec:Methods}).
The superconducting correlation at long distance and zero temperature
is accurately calculated.
The amplitude of the superconducting gap functions are \txo{also} estimated from
the single-particle momentum distribution.}

\txo{It is found that the in-plane superconducting correlation is not enhanced
in comparison with the superconducting correlation in the single-layer $t$-$t'$ Hubbard model.
While the superconducting correlations at long distance both in the single-layer and bilayer models are
quantitatively similar at the optimal doping where the superconducting correlation
becomes maximum,
the superconducting correlations of the bilayer hamiltonian are significantly small at
the larger doping region in comparison with the correlations of the single-layer hamiltonian.
The reduction at the overdoped region is attributed to the van Hove singularity \txo{of the non-interacting band structure}.
In the single-layer $t$-$t'$ Hubbard model, the superconducting gap opens across the van Hove singularity in the normal state,
in the wide range of the hole doping.
In contrast, the superconducting gap does not involve the van Hove singularity in the bilayer $t$-$t'$ Hubbard model at the overdoped region.
In addition, we
found that the amplitude of the superconducting gap functions are also
similar in both the single-layer and bilayer $t$-$t'$ Hubbard model at the optimal doping.
Thus, the adjacent CuO$_2$ layers are not relevant to the enhancement of $\Tc^{\rm opt}$ in the bilayer cuprates.
Other possible factors relevant to the $n$ dependence of $T_{\rm c}^{\rm opt}$
other than the adjacent layers are also discussed.}

{The organization of the present paper is as follows.
In Sec.~\ref{Sec:Preliminaries}, we review the previous studies on
superconductivity in single-layer and multilayer cuprates to make
the motivation of the present study.
\txo{Sections \ref{Sec:Model_and_Methods} and \ref{Sec:Methods} are} devoted to introducing
\txo{the} bilayer Hubbard-type hamiltonians and numerical methods used in the present study.
We show our results on the superconducting correlations \txo{and other physical quantities} of the bilayer systems in Sec.~\ref{Sec:Results}.
\txo{The summary of the present study and discussion on the results and these implications are given in Sec.~\ref{Sec:Summary_and_Discussion}}.}

\section{Preliminaries}\label{Sec:Preliminaries}

{In the present study, we
\txo{examined}
the impact of the adjacent CuO$_2$ layers
on the stability of the superconductivity in the multilayer cuprates.
To \txc{focus on} the impact, 
we \txc{\txo{studied} simple and relevant} effective hamiltonians \txc{to}
\txc{the single-layer} and bilayer cuprates.
\txc{To choose appropriate effective hamiltonians,}
we briefly summarize \txc{and examine} the previous results on \txc{single-layer} and multilayer cuprates
\txc{in the following section, with emphasis on theoretical and numerical studies}.
}

\subsection{{Single CuO$_2$ layer physics}}\label{sec:single_CuO2}
There have been numerous theoretical studies on \txo{properties of} a single CuO$_2$ layer.
\txo{At the early stage of the research, researchers got a consensus that}
{the electronic structure of the single CuO$_2$ layer around the Fermi level
is dominated by the antibonding band consisting of $d_{x^2-y^2}$ orbitals of Cu ions and 2$p$ orbitals of O ions
[the $d$-$p$ model (three-band model)~\cite{emery1987theory}].}
\tby{\txo{Afterwards}, the effective hamiltonians for the two-dimensional single-band system have been intensively studied.
The well-studied single-band effective hamiltonians
are
the $t$-$J$ model~\cite{PhysRevB.37.3759}
and Hubbard model~\cite{Kanamori,hubbard1963electron,Gutzwiller}
on square lattices.
The Hubbard model take into account both of the localized and itinerant nature of strongly correlated electrons
{while the $t$-$J$ model omits a part of the itinerant nature, namely, doublon formation}.}

The Hubbard model
is defined as,
\begin{eqnarray}
    H=\txo{-t\sum_{\langle i,j\rangle,\sigma}}c_{i\sigma}^{\dagger}c_{j\sigma}+U\sum_{i}n_{i\uparrow}n_{i\downarrow}, \label{eq:Hubbard}
\end{eqnarray}
where $c_{i\sigma}^{\dagger}$ ($c_{i\sigma}$) is a creation (annihilation) operator for
an electron at the $i$th site with spin $\sigma\ (=\uparrow, \downarrow)$, and $n_{i\sigma}=c_{i\sigma}^{\dagger}c_{i\sigma}$
is a particle number operator.
Here, \txo{$-t$ is the single-particle transfer integral or hopping
between $i$th and $j$th sites that constitute a pair of the nearest-neighbor sites,  
$\langle i,j\rangle$ denotes the pair of the nearest-neighbor sites,
and $U$ is the on-site Coulomb repulsion.}

There have been \tby{intensive} theoretical attempts to reveal the ground state of the Hubbard model \tby{on the square lattice}.
It has been believed that spatially uniform $d$-wave superconducting phases are stabilized
\tby{in the wide range of the hole doping~\cite{giamarchi1991phase,
capone2006competition,
yokoyama2012crossover,
misawa2014origin}}.
\tby{For example,} a variational Monte Carlo study revealed that
the phase separation between antiferromagnetic Mott insulators and superconducting states~\cite{misawa2014origin}
appears in the underdoped region.
{However simulations for larger system sizes revealed that} 
there are wide charge/spin stripe ordered phase and uniform $d$-wave SC phase was unstable~\txo{\cite{ido2018competition,darmawan2018stripe}},
which is consistent with other results obtained by different methods~\cite{zheng2017stripe}.

\txo{The next nearest-neighbor hopping $t'$ changes the instability towards the phase separation.
While the numerical study~\cite{aichhorn2007phase} by the variational cluster approach~\cite{PhysRevLett.91.206402}
shows the phase separation,
the phase separation
disappears in the previous mVMC studies
when the finite nearest-neighbor hopping, $t'/t=-0.3$, is introduced~\cite{misawa2014origin}.
The nature of the superconductivity is also altered by $t'$.
While, in the standard Hubbard model without $t'$,
the superconductivity coexists with the antiferromagnetic order,
it does not with finite $t'/t$~\cite{ido2018competition}.}

\tby{While the stripe orders become stable in the ground state,
the uniform superconductivity has been found in an eigenstate
of the 2D Hubbard model,
which
is found as a stable local minimum during the optimization of the variational wave function.
{In contrast to the phase diagram of the cuprate superconductors}\gr{~\cite{keimer2015quantum}},
{the uniform superconductivity in the Hubbard model is stabilized only in the overdoped region~\cite{darmawan2018stripe}. 
In addition, the superconductivity is too strong to explain experimental observations of the superconductivity in the cuprates.}}
\txo{The superconducting correlation function at the long distance (see \ref{Sec:Superconducting_correlation_function})
is optimally $0.04$ in the standard Hubbard model ($t'=0$) for $U/t=10$
~\cite{misawa2014origin,darmawan2018stripe}.
The superconducting gap in the Hubbard model is also estimated as $0.15t$ from
the single-particle spectral function~\cite{PhysRevX.10.041023}.}

{It has been revealed in the numerical study~\cite{ohgoe2020ab} based on
an {\it ab initio} hamiltonian derived for Hg-based cuprate superconductor~\cite{hirayama2018ab,hirayama2019effective}
that the long-range Coulomb repulsion relatively favors the uniform superconducting state in a wide doping range.
The discrepancy between the ground-state phase diagram of the Hubbard model and cuprate superconductors
is primarily attributed to the long-range Coulomb repulsion.}

{Although there are severe competition among several ordered states,
the uniform superconducting state \txc{is found to be} an eigenstate or a local minimum~\cite{ido2018competition,ohgoe2020ab}.
The strong superconducting order in the Hubbard model with the short-range interaction
is adiabatically connected to the reasonable superconducting order \txc{in the realistic hamiltonian with long-range Coulomb
repulsion, which is demonstrated for the {\it ab initio} hamiltonian of HgBa$_2$CuO$_{4+y}$~\cite{ohgoe2020ab}}.
\txo{The superconducting correlation is typically $0.005$ in the {\it ab initio} effective hamiltonian of
the Hg cuprate~\cite{ohgoe2020ab}, which is one order of magnitude smaller than the correlation in the Hubbard model.}}

\subsection{{Interlayer couplings}}



\txo{Here, we summarize
previous studies on
interlayer couplings
between adjacent CuO$_2$ layers.
There are considerable amount of studies on single-electron hoppings and tunnelings of a Cooper pair
among the adjacent CuO$_2$ layers.
Since hoppings of a pair of electrons do not require the formation of the Cooper pair,
the pair hoppings generated by interlayer Coulomb repulsion has also been studied.}
\txo{As reviewed below, the $n$ dependence of $T_{\rm c}^{\rm opt}$ is inconsistent with the stabilization of the superconductivity
due to the Cooper pair tunnelings.
Theoretical estimates of $T_{\rm c}^{\rm opt}$ by pair hoppings of electrons
have shown that the appropriate enhancement of $T_{\rm c}^{\rm opt}$ requires an amplitude of the pair hoppings
larger than those of the typical Hund's rule couplings.
The Cooper pair tunneling or pair hopping mechanism alone hardly
explain the quantitative enhancement of $T_{\rm c}^{\rm opt}$ due to the adjacent CuO$_2$ layers.
Therefore, in the present paper, we only take into account the interlayer single-electron hoppings
as an essential interlayer term
in low-energy effective hamiltonians of the multilayer cuprates.}

\subsubsection{\txo{Interlayer single-particle} hoppings} \label{sssec:bbs}
{The interlayer hoppings among the adjacent layers
have been studied by using spectroscopy.
Among multilayer cuprates,
a bilayer cuprate,
Bi2212, is the most intensively investigated cuprates by using 
angle-resolved photoemission spectroscopy (ARPES)
due to the availability of large high-quality single crystals,
and the presence of a natural cleavage plane between the BiO layers~\cite{damascelli2003angle}. 
For Bi2212,
one of characteristic closely related to SC is band splitting around antinodal point \txo{$(\pi/a,0)$} in Fermi surface,
\txo{where $a$ is the distance between the nearest-neighbor Cu ions in a CuO$_2$ plane,} i.e.,
there are two Fermi surface, the bonding band (BB) and antibonding band (AB).
Here, we ignore the small deformation in the non-tetragonal crystal structure of Bi2212.} 
In ARPES measurements, two Fermi surface are clearly observed around
\txc{the} antinodal point $(\pi/a,0)$ and converged at \txc{the} nodal line around $(\pi/2a,\pi/2a)$.

It was well confirmed that this electronic structure is due to the
\txo{single-electron hopping
between adjacent} $\mathrm{CuO_{2}}$ layers \cite{feng2001bilayer}. 
These Fermi surfaces are well consistent with the function form of
the interlayer single-particle hopping, $t_{\perp}[\cos(k_{x}a)-\cos(k_{y}a)]^2/2$,
which is obtained by {\it ab initio} electronic structure calculations~\cite{ANDERSEN19951573}.
\txo{Below, we often set $a=1$ for simplicity.}




\txo{Although simple nearest-neighbor single-particle or
momentum indepdendent hoppings between adjacent layers
have been examined in the literature~\cite{medhi2007coexistence,okamoto2008enhanced},
the function form $t_{\perp}[\cos(k_{x}a)-\cos(k_{y}a)]^2/2$ is employed in the present paper
to reproduce the decent bilayer splittings of the Fermi surfaces.
While the charge transfer among the CuO$_2$ layers may cause the self-doping~\cite{okamoto2008enhanced}
even in the bilayer system,
the self-doping was not found in the present study.
It has also been proposed that a substantial (momentum-independent) bilayer hopping
weakens the intralayer $d_{x^2\mathchar`-y^2}$-wave pairing
and promotes interlayer $s_{\pm}$-wave pairings~\cite{bulut1992nodeless,maier2011pair}.
Howerver, the bilayer hopping, $t_{\mathrm{bi}}$,
which is taken from Ref.~\citen{markiewicz2005one} and used in the present study,
is insufficient to stabilize the $s_{\pm}$-wave pairing.}

\subsubsection{Interlayer \txo{electron-pair tunnelings}}
{Instead of tunneling of a single electron,
tunneling of a Cooper pair shows another energy scale of
interlayer couplings in the superconducting phase.
There have been several proposals on the mechanism of the pair hoppings.}


One of \txo{these proposals} is \txo{the} interlayer tunneling theory (ILT) proposed by Chakravarty and Anderson~\cite{chakravarty1993interlayer}.
\txo{The} ILT explains
\txo{the enhancement of $\Tc$ in the multilayer cuprates is attributed to interlayer tunnelings} of Cooper pair via Josephson coupling
{arising \txo{through} a second order process of interlayer single-particle hopping}\txo{.}
{\txo{The} gain of kinetic energy along $c$-axis promotes the \txo{Cooper-pair} formation in the \txo{single CuO$_2$} plane according to the pair tunneling term.}
In the flamework of the ILT, $T_c(n)$ of the $n$-layer cuprate is \txo{a monotonically} increasing function of $n$: 
$T_{c}(n)=T_{c}(1)+C(1-1/n)$ 
where $C$ is a constant~\cite{chakravarty1998electrons}.
\txo{The $n$ dependence of $\Tc$ was also derived by taking into account
interlayer Coulomb repulsions~\cite{leggett1999cuprate}.}
{However, realistic \txo{energy scale} of the interlayer tunneling term
$t_{\mathrm{bi}}^2/t \sim 0.1$,
\txo{where $t_{\mathrm{bi}}$ is the interlayer hopping and $t_{\mathrm{bi}}/t\sim0.3$,}
is insufficient to enhance the critical temperatures significantly.}
Related to \txo{the} ILT,
Chakravarty also studied Josephson-like \txo{couplings}
between $\mathrm{CuO_{2}}$ layers by the \txo{phenomenological} Ginzburg-Landau theory~\cite{chakravarty2004explanation}.

The measurement of the $c$-axis optical response directly gives us the Josephson coupling energy.
The multilayer cuprates have more than two $\mathrm{CuO_{2}}$ layers in a unit cell, which means more than one kind of Josephson junctions, i.e. a bilayer cuprate is a stack of a stronger junction within a bilayer and and a weaker junction between bilayers.
In such structure, optical Josephson plasma modes appear like the optical phonon modes in a crystal with more than two inequivalent atoms in a unit cell \cite{uchida2014high}.
The strength of \txo{the} $c$-axis Josephson coupling is proportional to the square of the frequency of \txo{the} optical Josephson mode.
\txo{The $c$-axis} Josephson coupling within layers in multilayer cuprates is related to the enhancement of \txo{$\Tc^{\rm opt}$}.

The systematic study of \txo{the} Josephson plasma modes was carried out for Hg-based multilayer cuprates~\cite{hirata2012correlation},
which
shows the frequency of \txo{the} optical Josephson plasma modes vary with increasing \txo{the} number of \txo{the} $\mathrm{CuO_{2}}$ layers.
The \txo{$n$ dependence of the} Josephson coupling energy per layer calculated in Ref.~\citen{hirata2012correlation}
\txo{is consistent}
with \txo{the $n$ dependence of $\Tc^{\rm opt}$} that $\Tc^{\rm opt}$ rises for $n\leq 3$ and decreases for $n\geq4$
\txo{while the ILT is inconsistent with the $n$ dependence of $\Tc^{\rm opt}$}.

\txo{Another scenario is the hopping processes of
electron pairs, instead of the Cooper pairs,
arising from matrix elements of Coulomb interaction~\cite{doi:10.1143/JPSJ.78.114716}.}
\txo{The impacts of the pair hoppings on $\Tc$ were theoretically examined by using a weak coupling approach~\cite{nishiguchi2013superconductivity}.}
{In the weak coupling approach, the interlayer \txo{single-particle} hoppings do not explain the enhancement of $\Tc$.
Therefore, the authors of Ref.~\citen{nishiguchi2013superconductivity} attributed the enhancement to the interlayer pair hoppings.}
\txo{The amplitude of the pair hoppings
is required to be
comparable with $t$ to
explain the enhancement of $\Tc$.}
\txo{Although the pair hopping arising from matrix elements of the Coulomb repulsions~\cite{zegrodnik2017effect}
has been also examined by the Gutzwiller wave functions,
a significant enhancement of the gap function requires substantial amplitude of the pair hoppings comparable with $t$.}
\txo{In addition to the pair hoppings,
the interlayer exchange $J_{\perp}$, as another possible interlayer two-body interaction,
have been examined~\cite{medhi2007coexistence,okamoto2008enhanced,zegrodnik2017effect}.
It is highly desirable to perform quantitative and {\it ab initio} studies on whether
the interlayer pair hoppings and exchange couplings are enough large to explain the enhancement of $\Tc$, or not.}

\section{Model} \label{Sec:Model_and_Methods}
\txc{In contrast to these previous study, present study aims to investigate the rise of $\Tc$ in multilayer cuprates from microscopic perspective by using numerical method beyond mean-field approximations and weak coupling approaches.}
\txc{In this study, we
investigate the stabilization of SC in the multilayer cuprates by
a well-tested numerical method.}
It is necessary to examine microscopically how the property of SC is varied by the multilayer effect.
As a first step, we examine the ground state of bilayer Hubbard model with interlayer single-particle hopping, which is considered to be the most fundamental model for multilayer cuprates.
For the better understanding of the SC in the bilayer cuprates, we focus on pairing structure or correlation which includes the interlayer SC correlation as well as the intralayer correlation.

\subsection{Bilayer $t$-$t'$ Hubbard model}
\txc{In this paper, we focus on an isolated bilayer
and study}
\txc{the following} bilayer Hubbard model,
\begin{eqnarray}
    \mathcal{H}_{\mathrm{bi}}=-\sum_{i,j=1}^{\Ns}\sum_{\alpha,\beta=1,2}\sum_{\sigma}t_{ij}^{\alpha\beta}
{c_{i\sigma}^{\alpha}}^{\dagger} c_{j\sigma}^{\beta} +U\sum_{i,\alpha}n_{i\uparrow}^{\alpha}n_{i\downarrow}^{\alpha}, \label{eq:bilayerHubbard}
\end{eqnarray}
where 
\txc{$\alpha,\beta$ are the layer indices,
${c_{i\sigma}^{\alpha}}^{\dagger}$
($c_{i\sigma}^{\alpha}$) is the creation (annihilation)
operator that generates (destroys) the $\sigma$ spin electron
at the $i$th site of the $\alpha$th layer,
$n_{i\sigma}^{\alpha}={c_{i\sigma}^{\alpha}}^{\dagger}c_{i\sigma}^{\alpha}$,}
and {$\Ns=L\times L$} is the number of sites per layer.
\txc{The first term in the right hand side of Eq.~(\ref{eq:bilayerHubbard})
is kinetic energy $\mathcal{H}_t$,
which is rewritten in momentum space as follows,}
\begin{eqnarray}
    \mathcal{H}_{t}=\sum_{\bm{k}\sigma}
\left(
{c_{\bm{k}\sigma}^{1}}^{\dagger}\ {c_{\bm{k}\sigma}^{2}}^{\dagger}
\right)
    \left(\begin{array}{cc}
    \epsilon_{\bm{k}} & t_{\bm{k}}\\
    t_{\bm{k}} & \epsilon_{\bm{k}}
    \end{array}\right)
    \left(\begin{array}{c} c_{\bm{k}\sigma}^{1}\\ c_{\bm{k}\sigma}^{2} \end{array}\right),
\label{eq:bi_tight-binding}
\end{eqnarray}
\txc{where $\bm{k}$ is the in-plane momentum, and} $c_{\bm{k}\sigma}^{\alpha}$ is the Fourier transform of $c_{i\sigma}^{\alpha}$:
\begin{eqnarray}
    c_{\bm{k}\sigma}^{\alpha}=\frac{1}{\sqrt{\Ns}}\sum_{i=1}^{\Ns} e^{-i\bm{k}\cdot\bm{r}_{i}}c_{i\sigma}^{\alpha}. \label{eq:fourier_def}
\end{eqnarray}
\txc{Here, $\epsilon_{\bm{k}}$ is the intralayer energy dispersion and $t_{\bm{k}}$ is the interlayer hybridization,
which can be chosen to be real.} 

\begin{figure}[htbp]
\begin{center}
    \includegraphics[width=\linewidth]{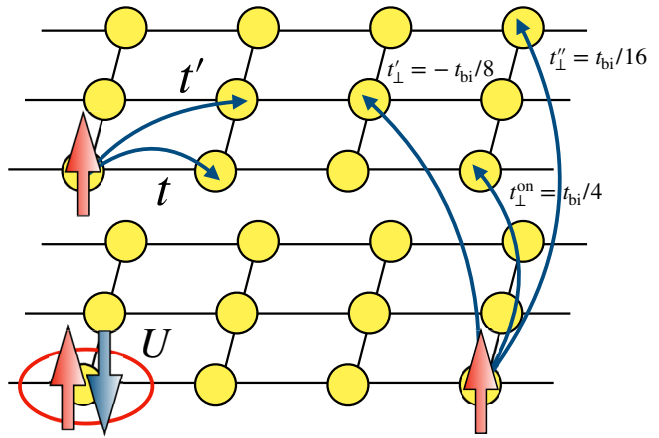}
\end{center}
\caption{\txc{Bilayer Hubbard hamiltonian studied in the present paper.
The nearest-neighbor and second-nearest neighbor intralayer hoppings are
represented by $t$ and $t'$, respectively. 
The on-site Coulomb repulsion is denoted by $U$.
The nearest-neighbor, third-nearest-neighbor, and fourth-nearest-neighbor interlayer hoppings
are denoted by {$t_{\perp}^{\rm on}=t_{\rm bi}/4$, $t_{\perp}'=-t_{\rm bi}/8$, and $t_{\perp}''=t_{\rm bi}/16$},
respectively [see Eq.~(\ref{eq:bilayerHop})].}
}
    \label{fig:bilayerHubbard}
\end{figure}

\txc{To take essential physics of the {antibonding band} in each CuO$_2$ layer,
we introduce the nearest-neighbor and next-nearest-neighbor intralayer hoppings, $t$ and $t'$, respectively. 
Then, the intralayer energy dispersion is given by
\begin{eqnarray}
    \epsilon_{\bm{k}}=-2 t\left(\cos{k_{x}}+\cos{ k_{y}}\right)-4 t^{\prime} \cos{k_{x}}\cos{k_{y}}
\label{eq:intralayer_eps}
\end{eqnarray}}
By following the literature~\cite{ANDERSEN19951573,markiewicz2005one}, 
we choose the following interlayer term, 
\begin{eqnarray}
    t_{\bm{k}}=-\frac{t_{\mathrm{bi}}}{4}(\cos{k_x}-\cos{k_y})^2, \label{eq:bilayerHop}
\end{eqnarray}
which is originally proposed by Chakravarty~\cite{chakravarty1993interlayer},
and later confirmed by
\txc{derivation of the low-energy hamiltonians based on the local density approximation (LDA)}~\cite{ANDERSEN19951573}.
\txc{When we introduce the nearest-neighbor interlayer hopping, $t_{\perp}^{\rm on}=t_{\rm bi}/4$,
third-nearest-neighbor interlayer hopping, $t_{\perp}'=-t_{\rm bi}/8$,
and fourth-nearest-neighbor interlayer hopping, $t_{\perp}''=t_{\rm bi}/16$ {(see Fig.~\ref{fig:bilayerHubbard})},
the interlayer term $t_{\bm{k}}$ is given by Eq.~(\ref{eq:bilayerHop}).}
\txc{In Ref.~\citen{ANDERSEN19951573}, the interlayer hopping $t_{\bm{k}}$
is derived for YBa$_2$Cu$_3$O$_7$ where a Y layer is sandwiched by two adjacent CuO$_2$ layers,
while the same momentum dependence of the interlayer hopping is shown
in Bi2212 where a Ca layer is sandwiched by the CuO$_2$ layers~\cite{markiewicz2005one}.}

\txc{We determine the hoppings, $t$, $t'$, and $t_{\rm bi}$
by following
the tight-binding fitting to the LDA results~\cite{markiewicz2005one}.
In Ref.~\citen{markiewicz2005one},
$t=360$ meV, $t'=-100$ meV, and $t_{\mathrm{bi}}=110$ meV
are estimated for Bi2212.
Therefore, we use $t'/t = -100/360$ $(\simeq -0.28)$ and $t_{\mathrm{bi}}/t = 110/360$ $(\simeq 0.3)$}.
\txc{The on-site Coulomb repulsion $U$ is estimated to be around 4 eV
for the antibonding $d_{x^2\mathchar`-y^2}$ orbital of the cuprates~\cite{hirayama2019effective}.
Thus, we choose $U/t=10$
as a typical value.} 

\begin{table}[htb]
\begin{center}
\caption{\txc{Elements of the hopping matrix and amplitude of the Coulomb repulsion
of the bilayer $t$-$t'$ Hubbard model}.
\label{table:hopping}}
\begin{tabular}{ccccc}
    &&&&\\
\hline
\hline
$t'/t$ && $t_{\rm bi}/t$ && $U/t$ \\
\hline
$-100/360$ &&
$110/360$ && 10 \\
\hline
\hline
\end{tabular}
\label{table_gap}
\end{center}
\end{table}

\subsection{Bonding and antibonding band}
\begin{figure}[htbp]
    \centering
    \includegraphics[width=0.3\textwidth]{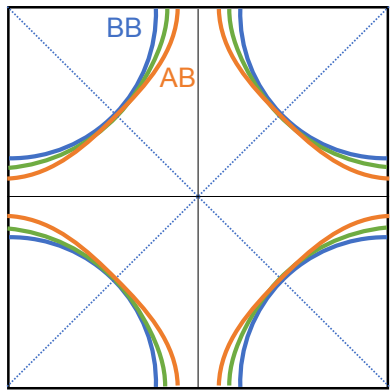}
    \caption{\txc{Non-interacting Fermi surfaces of BB and AB at} \gr{the half-filling.
\txo{The blue}, red, and green curves show \txo{the BB, AB, and single-layer Fermi surface,} respectively.}}
    \label{fig:bi_bandsplit}
\end{figure}

As mentioned in Sec.~\ref{sssec:bbs}, for bilayer cuprates such as Bi2212,
\txc{the bonding band (BB) and antibonding band (AB) are observed in the momentum space. 
By diagonalizing the tight-binding hamiltonian Eq.~(\ref{eq:bi_tight-binding}),
we can reproduce the band splitting between BB and AB.} 
The diagonalized tight-binding hamiltonian is 
\begin{eqnarray}
    \mathcal{H}_{t}&=&
    \sum_{k\sigma}
\left({c_{\bm{k}\sigma}^{+}}^{\dagger}\ {c_{\bm{k}\sigma}^{-}}^{\dagger}\right)
\left(\begin{array}{cc}
\epsilon_{\bm{k}}^{+}
& 0 \\ 0 &
 \epsilon_{\bm{k}}^{-}
\end{array}\right)
\left(\begin{array}{c} c_{\bm{k}\sigma}^{+}\\ c_{\bm{k}\sigma}^{-} \end{array}\right),
\nonumber\\
\label{eq:bi_tight_diag}
\end{eqnarray}
where
\txcy{$\epsilon_{\bm{k}}^{+} = \epsilon_{\bm{k}}+t_{\bm{k}}$
($\epsilon_{\bm{k}}^{-} = \epsilon_{\bm{k}}-t_{\bm{k}}$)
is the BB (AB) band dispersion.}
\txc{Here,
${c_{\bm{k}\sigma}^{+}}^{\dagger}$
and
${c_{\bm{k}\sigma}^{+}}$
(${c_{\bm{k}\sigma}^{-}}^{\dagger}$
and
${c_{\bm{k}\sigma}^{-}}$)
are the creation and annihilation operators of $\sigma$ spin quasiparticle in BB (AB), respectively. 
These} fermion operators $c_{\bm{k}\sigma}^{\pm}$ are given by 
\begin{eqnarray}
    c_{\bm{k}\sigma}^{\pm}=\frac{1}{\sqrt{2}}(c_{\bm{k}\sigma}^{1}\pm c_{\bm{k}\sigma}^{2}).
\end{eqnarray}
\txc{When} $t_{\bm{k}}$ is \txc{taken} as Eq.~\ref{eq:bilayerHop},
the non-interacting Fermi \txc{surfaces at the half-filling} are shown in Fig.~\ref{fig:bi_bandsplit}.
\txc{Due to the momentum dependence of $t_{\bm k}$, BB and AB are degenerated along the nodal line that connects
${\bm k}=(0,0)$ and $(\pi,\pi)$ while they shows the splitting around the antinodal region.}

\section{Methods}
\label{Sec:Methods}
{In this study,
\txc{a highly flexible variational Monte Carlo method (VMC) is utilized to obtain the ground state wave function.}
\txc{To perform the VMC simulation,} we used an open-source software package,
\txc{many-variable} variational Monte Carlo method (mVMC)~\cite{tahara2008variational,MISAWA2019447}.}

\subsection{Variational wave function}
\txc{In the present study, we introduce the following variational wave functions,}
\begin{eqnarray}
    \ket{\psi}=\mathcal{P}_{\mathrm{G}}\mathcal{\mathcal{P}}_{J}\mathcal{P}_{\mathrm{d\mathchar`-h}}^{\mathrm{ex}}
\txo{
\mathcal{L}^{S}}
\ket{\phi_{\mathrm{pair}}},
\end{eqnarray}
\txcy{for the single-layer system, and,}
\begin{eqnarray}
\gr{\ket{\psi}}=\mathcal{P}_{\mathrm{G}}\mathcal{\mathcal{P}}_{J}\mathcal{P}_{\mathrm{d\mathchar`-h}}^{\mathrm{ex}}\ket{\phi_{\mathrm{pair}}},
\label{eq:wf}
\end{eqnarray}
\txcy{for the bilayer system,}
\txc{where $\ket{\phi_{\mathrm{pair}}}$ is a pair-product wave function,
$\mathcal{P}_{\mathrm{G}}, \mathcal{\mathcal{P}}_{J}$, and $\mathcal{P}_{\mathrm{d-h}}^{\mathrm{ex}}$ are
the Gutzwiller~\cite{gutzwiller1963effect}, Jastrow~\cite{jastrow1955many}, and doulon-holon~\cite{yokoyama1990variational}
correlation factors, respectively,
and
\txo{$\mathcal{L}^{S}$ 
is the spin quantum-number projection}~\cite{PhysRevB.69.125110,tahara2008variational}.}
\txcy{As explained below,
any Hartree-Fock-Bogoliubov-type wave function
is represented by the pair-product wave function.}
\gr{Here, we \txcy{do not employ} the spin quantum-number projection $\mathcal{L}^{S}$ for the present study of the bilayer $t$-$t'$ Hamiltonian
\txcy{to save the computational resources.}
Although the spin quantum-number projection improves the ground state energy,
the superconducting correlation is \txcy{not
affected} by the spin quantum-number projection\txcy{, as demonstrated in Appendix~\ref{chap:SPvsNP} for the bilayer system.}}

\subsubsection{Pair-product state}
\txc{The pair-product wave function $\ket{\phi_{\mathrm{pair}}}$ is
defined as}
\begin{eqnarray}
    \ket{\phi_{\mathrm{pair}}}=
\left(\sum_{i,j=1}^{\Ns}\sum_{\alpha,\beta=1,2}f_{ij}^{\alpha\beta}
{c_{i\uparrow}^{\alpha}}^{\dagger}
{c_{j\downarrow}^{\beta}}^{\dagger}\right)^{N_{\mathrm{e}}/2}\ket{0},
\end{eqnarray}
where $f_{ij}^{\alpha\beta}$
\txc{is a variational parameter, $N_{\rm e}$ is the number of the electrons,
and $\ket{0}$ is a vacuum.}
\txc{Although we could optimize $(2\Ns)^2$ variational parameters, $f_{ij}^{\alpha\beta}$,
we reduce the number of independent variational parameters by partially imposing
translational symmetry on $f_{ij}^{\alpha\beta}$.}
\txc{Here, we impose a $2\times 2$ sublattice structure or a $2\times 2$ supercell,
and assume that the wave function is invariant under the translations {(2a, 0)
and (0, 2a).}
Since there are two orbitals or layer degrees of freedom at each site,
there are $2^3$ orbitals in the supercell.}
Then, due to the translational symmetry, 
there are \txc{$2^3 \times 2^3 \times (\Ns / 2^2)$} independent variational parameters
for the
\txc{pair-product wave function.}

\subsubsection{Correlation factors}

\txc{The Gutzwiller factor~\cite{gutzwiller1963effect} controls
the number of the doubly occupied sites
through the variational parameters $g_{i}^{\alpha}$ defined at each site as below},
\begin{eqnarray}
    \mathcal{P}_{\mathrm{G}}=\exp\left(-\sum_{i,\alpha}g_{i}^{\alpha}n_{i\uparrow}^{\alpha}n_{i\downarrow}^{\alpha}\right) .
\end{eqnarray}
In the limit of $g_{i}^{\alpha}\rightarrow\infty$,
the $\mathcal{P}_{\mathrm{G}}$ totally \txc{excludes} the {double occupation}.
\txc{In the present study,}
\txc{the sublattice periodicity
is also imposed on 
the parameter $g_{i}^{\alpha}$.}
\txc{Thus}, the number of \txc{the independent variational} parameters $g_{i}^{\alpha}$ is \txc{$2^3$}.

The Jastrow factor~\cite{jastrow1955many}
\txc{introduces long-range charge-charge correlations, which is defined as,}
\begin{eqnarray}
    \mathcal{P}_{\mathrm{J}}=\exp\left(-\frac{1}{2}\sum_{i,\alpha,j,\beta}v_{ij}^{\alpha\beta}n_{i}^{\alpha}n_{j}^{\beta}\right).
\end{eqnarray}
\txc{Here, we set $v_{ij}^{\alpha\beta}=0$ for $i=j$ and $\alpha=\beta$
since the on-site correlation is already introduced by $\mathcal{P}_{\mathrm{G}}$.
We also assume the $2\times 2\times 2$ sublattice structure of $v_{ij}^{\alpha\beta}$.}

The doublon-holon factor~\cite{yokoyama1990variational} is defined as
\begin{eqnarray}
    \mathcal{P}_{\mathrm{d}\mathchar`-\mathrm{h}}^{\mathrm{ex}}=\exp \left[-\sum_{m=0}^{4} \sum_{\ell=1,2} \alpha_{(m)}^{(\ell)} \sum_{i} \xi_{i(m)}^{(\ell)}\right],
\end{eqnarray}
\txc{where $\alpha_{(m)}^{(\ell)}$ is a variational parameter.}
\txc{Here, $\xi_{i(m)}^{(\ell)}$ is a many-body operator that is diagonal in the real-space
electron configurations and is given by Ref.~\citen{tahara2008variational} as follows:
$\xi_{i(m)}^{(\ell)}=1$ if a doublon (holon) exists at the $i$th site and
is surrounded by $m$
holons
(doublons)
at the $\ell$th nearest neighbor. Otherwise, $\xi_{i(m)}^{(\ell)}=0$.} 

\subsubsection{Initial wave functions}
\txc{Even though the variational wave function Eq.~(\ref{eq:wf}) is designed to be highly flexible,
the choice of the initial guess for the variational parameters matters to the optimized wave function.}
\txc{When, for example, we examine whether the superconducting state is stable or not,
we prepare
a $d$-wave superconducting mean-field wave function as an initial guess}.

\txc{To obtain a mean-field superconducting state,
we introduce a mean-field BCS hamiltonian for bilayer lattice,
which is represented by the creation (annihilation) operators of the BB/AB band
${c_{\bm{k}\sigma}^{\pm}}^{\dagger}$
(${c_{\bm{k}\sigma}^{\pm}}$)
as}
\begin{eqnarray}
 H_{\mathrm{MF}} &=& \sum_{\bm{k},\sigma}\left[
\epsilon_{\bm{k}}^{+}{c_{\bm{k}\sigma}^{+}}^{\dagger}c_{\bm{k}\sigma}^{+}
+\epsilon_{\bm{k}}^{-}{c_{\bm{k}\sigma}^{-}}^{\dagger}c_{\bm{k}\sigma}^{-}\right. \non\\
&& +\Delta_{\mathrm{SC}}^{+}(\bm{k})
({c_{\bm{k}\uparrow}^{+}}^{\dagger}{c_{-\bm{k}\downarrow}^{+}}^{\dagger}+c_{-\bm{k}\downarrow}^{+}c_{\bm{k}\uparrow}^{+}) \non\\
&& \left. +\Delta_{\mathrm{SC}}^{-}(\bm{k})({c_{\bm{k}\uparrow}^{-}}^{\dagger}{c_{-\bm{k}\downarrow}^{-}}^{\dagger}+c_{-\bm{k}\downarrow}^{-}c_{\bm{k}\uparrow}^{-})\right] \non\\
&& -\mu_0\sum_{i,\alpha,\sigma}{c_{i\sigma}^{\alpha}}^{\dagger}c_{i\sigma}^{\alpha},
\end{eqnarray}
where $\epsilon_{\bm{k}}^{\pm}=\epsilon_{\bm{k}}\pm t_{\bm{k}}$
\txc{[see Eq.~(\ref{eq:bi_tight_diag})]} and $\mu_0$ is a chemical potential.
An eigenstate of $H_{\mathrm{MF}}$ is given by
\begin{eqnarray}
\ket{\phi_{\mathrm{SC}}}
&=&
\prod_{\bm{k}}\left[
\left(u_{\bm{k}}^{+}+v_{\bm{k}}^{+}{c_{\bm{k}\uparrow}^{+}}^{\dagger}{c_{-\bm{k}\downarrow}^{+}}^{\dagger}\right)
\right.
\non\\
&&\times
\left.
\left(u_{\bm{k}}^{-}+v_{\bm{k}}^{-}{c_{\bm{k}\uparrow}^{-}}^{\dagger}{c_{-\bm{k}\downarrow}^{-}}^{\dagger}\right)
\right]
\ket{0},
\end{eqnarray}
where
\begin{eqnarray}
u_{\bm{k}}^{\pm}
&=&
\frac{1}{\sqrt{2}}
\left\{
1+\frac{\xi_{\bm{k}}^{\pm}}{\sqrt{\left(\xi_{\bm{k}}^{\pm}\right)^2+\left[\Delta_{\mathrm{SC}}^{\pm}(\bm{k})\right]^2}}
\right\}^{1/2}, \\
v_{\bm{k}}^{\pm}
&=&
\frac{1}{\sqrt{2}}
\left\{
1-\frac{\xi_{\bm{k}}^{\pm}}{\sqrt{\left(\xi_{\bm{k}}^{\pm}\right)^2+\left[\Delta_{\mathrm{SC}}^{\pm}(\bm{k})\right]^2}}
\right\}^{1/2},\label{eq:v_BCS}
\end{eqnarray}
and
\begin{eqnarray}
\xi_{\bm{k}}^{\pm}=\epsilon_{\bm{k}}^{\pm}-\mu_0.
\end{eqnarray}

\txc{While the variational wave function $\ket{\psi}$ [Eq.~(\ref{eq:wf})] is an eigenstate of the electron number,
\eqsa{
\txc{\hat{N}=\sum_{i}\sum_{\alpha=1,2}\sum_{\sigma=\uparrow,\downarrow}{c_{i\sigma}^{\alpha}}^{\dagger}c_{i\sigma}^{\alpha}},
} 
the mean-field wave function $\ket{\phi_{\mathrm{SC}}}$ is not an eigenstate of $\hat{N}$.}
Then, to make an initial guess for $\ket{\psi}$,
we extract the $N_{\mathrm{e}}$-electron sector of $\ket{\phi_{\mathrm{SC}}}$ as,
\begin{eqnarray}
\ket{\phi_{\mathrm{SC}}^{N_{\mathrm{e}}}}=
\left[\sum_{\bm{k}}(g_{\bm{k}}^{+}{c_{\bm{k}\uparrow}^{+}}^{\dagger}{c_{-\bm{k}\downarrow}^{+}}^{\dagger}
+g_{\bm{k}}^{-}{c_{\bm{k}\uparrow}^{-}}^{\dagger}{c_{-\bm{k}\downarrow}^{-}}^{\dagger})\right]^{N_{\mathrm{e}}/2}\ket{0},
\end{eqnarray}
where
\begin{eqnarray}
 g_{\bm{k}}^{\pm}=\frac{v_{k}^{\pm}}{u_{k}^{\pm}}
=\frac{\Delta_{\mathrm{SC}}^{\pm}(\bm{k})}{\xi_{\bm{k}}^{\pm}+\sqrt{\left(\xi_{\bm{k}}^{\pm}\right)^2+
\left[\Delta_{\mathrm{SC}}^{\pm}(\bm{k})\right]^2}}.
\end{eqnarray}
Using the Fourier transformation Eq.~(\ref{eq:fourier_def}),
\txc{the pair-product wave function equivalent to $\ket{\phi_{\mathrm{SC}}}$ is obtained as,}
\begin{eqnarray}
 \ket{\phi_{\mathrm{SC}}^{N_{\mathrm{e}}}}&=\displaystyle\left[\sum_{i,\alpha,j,\beta}
f_{ij}^{\alpha\beta}{c_{i\uparrow}^{\alpha}}^{\dagger}{c_{j\downarrow}^{\beta}}^{\dagger}\right]^{N_{\mathrm{e}}/2}\ket{0},\\
 f_{ij}^{\alpha\beta} &=\displaystyle \frac{1}{2N_s}\sum_{\bm{k}}e^{i\bm{k}\cdot(\bm{r}_i-\bm{r}_j)}g_{\bm{k}}^{\alpha\beta},\\
 g_{k}^{\alpha\beta} &= \left\{\begin{array}{c}
			 g_{\bm{k}}^{+}+g_{\bm{k}}^{-} \quad(\alpha=\beta)\\
			 g_{\bm{k}}^{+}-g_{\bm{k}}^{-} \quad(\alpha\neq\beta)
			       \end{array} \right..
\end{eqnarray}

\txc{In the present paper,
we assume that $\Delta_{\mathrm{SC}}^{\pm}(\bm{k})$ has $d_{x^2-y^2}$-wave symmetry
and the simplest form as
$\Delta_{\mathrm{SC}}^{\pm}(\bm{k})=\Delta_{d}(\cos{k_{x}}-\cos{k_{y}})$.}
\txc{To prepare the initial guesses for the following simulation,
we choose the gap function depending on the doping in the range of $0.1 \leq \Delta_{d}/t \leq 0.5$.} 
\txc{The non-interacting Fermi energy is taken as the chemical potential $\mu_0$}.

\subsection{Optimization method}
\txc{All the parameters are optimized by
minimizing the energy expectation value,
\eqsa{
E_{\bm{\alpha}}
=
\frac{\bra{\psi_{\bm{\alpha}}}
\mathcal{H}_{\rm bi}
\ket{\psi_{\bm{\alpha}}}}
{\left\langle \psi_{\bm{\alpha}} \right| \left. \psi_{\bm{\alpha}} \right\rangle},
}
where $\bm{\alpha}$ is the set of the variational parameters
and the $\bm{\alpha}$ dependence of the variational wave function is explicitly denoted
by $\ket{\psi_{\bm{\alpha}}}$ instead of $\ket{\psi}$.
The optimization of the variational wave function
is performed by using the stochastic reconfiguration (SR) method~\cite{sorella2001generalized},
which is the imaginary time evolution projected onto the subspace spanned by
the category of the variational wave functions given in Eq.~(\ref{eq:wf})}~\cite{McLachlan1964,neuscamman2012optimizing}.
\txc{The SR method is essentially equivalent to the natural gradient~\cite{Amari1998}, which is one of the standard
optimization methods in neural network and machine learning community.}
\txc{The implementation of the SR method in mVMC
is detailed in Refs.~\citen{tahara2008variational} and \citen{MISAWA2019447}}.

\subsection{\txc{Observables}}
To investigate the ground state of \txc{the bilayer $t$-$t'$ Hubbard model,
we evaluate the expectation values of static correlation fuctions as,
$\bra{\psi}{c_{i\sigma}^{\alpha}}^{\dagger}c_{j\tau}^{\beta}\ket{\psi}/
\left\langle \psi \right|\left. \psi\right\rangle$
and
$\bra{\psi}
{c_{i\sigma}^{\alpha}}^{\dagger}
c_{j\tau}^{\beta}
{c_{k\lambda}^{\gamma}}^{\dagger}
c_{\ell\nu}^{\delta}
\ket{\psi}/
\left\langle \psi \right|\left. \psi\right\rangle$,
where $\sigma, \tau, \lambda$ and $\nu$ are spin indices, and
$\alpha, \beta, \gamma$ and $\delta$ are layer indices.}
\txc{Here, we focus on} the spin structure factors,
the intralayer $d$-wave superconducting correlations,
and
the momentum distribution function. 

\subsubsection{Spin structure factor}
The intralayer spin structure factor is defined by 
\begin{eqnarray}
    S^{\alpha}(\bm{q})=
{\frac{1}{N_{\rm s}}} \sum_{i, j}
\frac{
\bra{\psi}
\bm{S}_{i}^{\alpha} \cdot \bm{S}_{j}^{\alpha}
\ket{\psi}
}{\left\langle \psi \right|\left. \psi\right\rangle}
e^{\mathrm{i} \bm{q} \cdot\left(\bm{r}_{i}-\bm{r}_{j}\right)}. \label{eq:spin_cor}
\end{eqnarray}
\txc{Here,} $\bm{S}_{i}^{\alpha}$ is a local spin operator \txc{defined by}
\begin{eqnarray}
    \bm{S}_{i}^{\alpha}=\frac{1}{2}\sum_{\sigma,\sigma'}
{c_{i\sigma}^{\alpha}}^{\dagger}\bm{\sigma}_{\sigma\sigma'}c_{i\sigma'}^{\alpha},
\end{eqnarray}
\txc{where $\bm{\sigma}=(\sigma_x,\sigma_y,\sigma_z)$ is the vector consisting of the Pauli matrices,
$\sigma_x$, $\sigma_y$, and $\sigma_z$.}

\subsubsection{Superconducting correlation function}
\label{Sec:Superconducting_correlation_function}
The intralayer superconducting correlation function is defined as
\begin{eqnarray}
    P_{\mathfrak{s}}^{\alpha}(\bm{r})
&=&\frac{1}{2N_s}\sum_{i=1}^{\Ns}
\left(
\bra{\psi} 
{\Delta_{\mathfrak{s}}^{\alpha}}^{\dagger}(\bm{r}_i)
\Delta_{\mathfrak{s}}^{\alpha}(\bm{r}_i+\bm{r})
\ket{\psi}
/\left\langle \psi \right|\left. \psi\right\rangle
\right.
\non\\
&&
+
\left.
\bra{\psi} 
\Delta_{\mathfrak{s}}^{\alpha}(\bm{r}_i)
{\Delta_{\mathfrak{s}}^{\alpha}}^{\dagger}(\bm{r}_i+\bm{r})
\ket{\psi}
/\left\langle \psi \right|\left. \psi\right\rangle
\right),
\label{eq:pair_cor}
\end{eqnarray}
\txc{where the index $\mathfrak{s}$
denotes
the symmetry of the Cooper pair, and}
\txc{the singlet
pairing operator $\Delta_{\mathfrak{s}}^{\alpha}(\bm{r}_i)$ is} defined as
\begin{eqnarray}
    \Delta_{\mathfrak{s}}^{\alpha}(\bm{r}_i)
=\frac{1}{\sqrt{2}}\sum_{\bm{r}}f_{\mathfrak{s}}(\bm{r})(c_{\bm{r}_i\uparrow}^{\alpha}c_{\bm{r}_i+\bm{r}\downarrow}^{\alpha}
 -c_{\bm{r}_i\downarrow}^{\alpha}c_{\bm{r}_i+\bm{r}\uparrow}^{\alpha}),
\label{eq:Delta_s}
\end{eqnarray}
\txc{where $f_{\mathfrak{s}}(\bm{r})$ is the form factor of the Cooper pair.}
\txc{For a simple $d_{x^2\mathchar`-y^2}$-wave SC ($\mathfrak{s}=d_{x^2\mathchar`-y^2}$), the form factor is assumed as} 
\begin{eqnarray}
    f_{d_{x^2-y^2}}(\bm{r})=\delta_{r_y,0}(\delta_{r_x,1}+\delta_{r_x,-1})-\delta_{r_x,0}(\delta_{r_y,1}+\delta_{r_y,-1}).
\non\\
\label{eq:form_factor}
\end{eqnarray}
To evaluate the long-range part of \txc{intralayer SC correlations},
we average the SC correlations for $r_{\mathrm{min}}<r=|\bm{r}|<r_{\mathrm{max}}$ as
\begin{eqnarray}
    \bar{P}_{\mathfrak{s}}^{\alpha}
=\frac{1}{M}\sum_{r_{\mathrm{min}}<r=|\bm{r}|<r_{\mathrm{max}}}P_{\mathfrak{s}}^{\alpha}(\bm{r}),\label{eq:pair_cor_ave}
\end{eqnarray}
where $M$ is the number of \txo{the lattice point that satisfies} \txcy{$r_{\mathrm{min}}<r<r_{\mathrm{max}}$}.
\gr{\txcy{Here,} we set the lower limit $r_{\mathrm{min}}$ to $L/2\sqrt{2}$.}
\txc{Since we take the (anti-)periodic boundary condition,
we set the upper limit $r_{\mathrm{max}}$ to $L/\sqrt{2}$.}

\subsubsection{Momentum distribution function}

\txo{To access information of single-particle dispersion and superconducting gap functions,
simulations of single-particle spectra are straight forward.
However, the simulation costs much more than the ground-state simulations.
The single-particle momentum distribution function is
an alternative approach to such information.}

\txc{There is a choice of the Wannier orbitals to evaluate the momentum distribution function.
When the superconducting gap
at the Fermi surface 
is a major concern,
the momentum distribution function for the BB/AB orbital,}
\begin{eqnarray}
n_{\bm{k}}^{\pm}=
\txcy{\frac{1}{2}}\sum_{\sigma}
\frac{
\bra{\psi}
{c_{\bm{k}\sigma}^{\pm}}^{\dagger}
c_{\bm{k}\sigma}^{\pm}
\ket{\psi}
}
{\left\langle \psi \right|\left. \psi\right\rangle},
\end{eqnarray} 
\txc{will be relevant to superconductivity.}

\begin{figure}[htbp]
    \centering
    \includegraphics[width=0.45\textwidth]{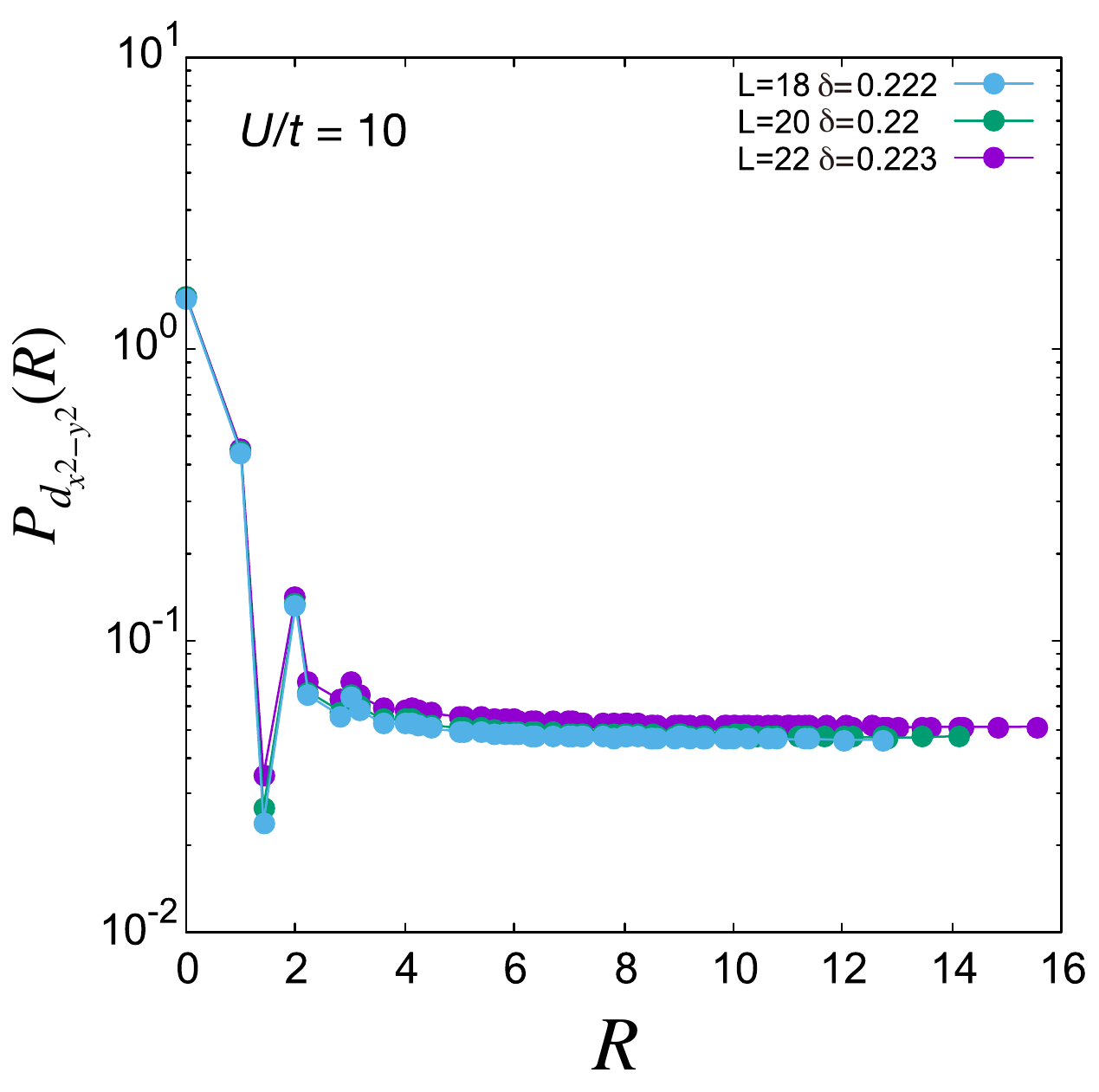}
    \caption{
\txc{Real-space superconducting correlation in the bilayer $t$-$t'$ Hubbard hamiltonian at $\delta \sim 0.22$.
{The system size dependence of $P_{d_{x^2\mathchar`-y^2}}^{\alpha}$ is examined for $L=18$,
$20$, and $22$ with the PP boundary condition}.}}
    \label{fig:superconducting_correlations}
\end{figure}

\begin{figure*}[tb]
        \centering
        \includegraphics[width=0.8\linewidth]{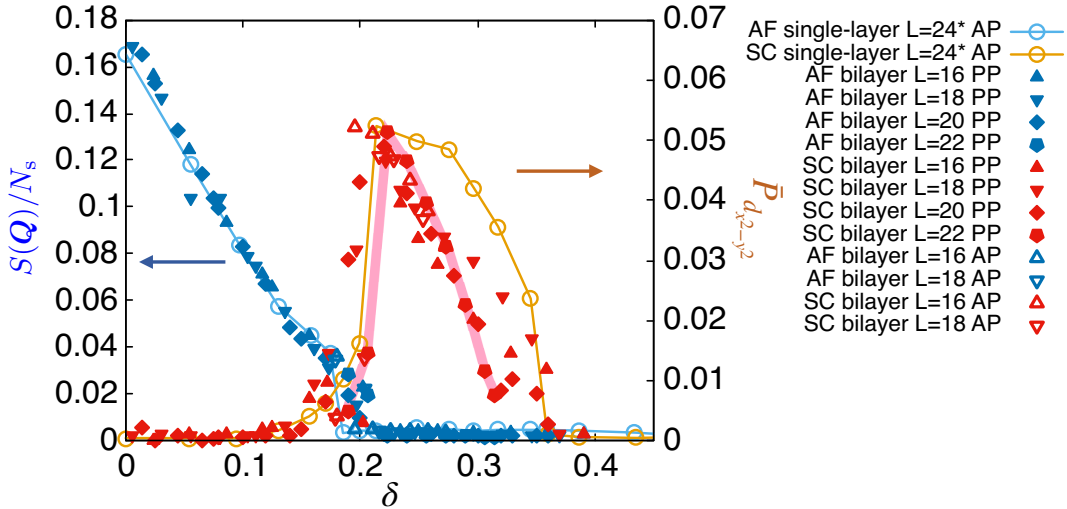}
        \caption{\txc{Doping dependence of spin structure factors and superconducting correlations at long distance
for the bilayer $t$-$t'$ Hubbard hamiltonian for
$t'/t=-100/360$, $t_{\rm bi}/t=110/360$, and $U/t=10$.}
{Doping rate $\delta$ is defined as $\delta=1-N/\Ns$. 
In present study, the number of sites per layer is $\Ns=L\times L$ with $L=16, 18, 20$, and 22. 
\txo{The blue} symbols (AF) show the peak \txcy{amplitude} of spin structure factors $S(Q)$ and red symbols (SC) show superconducting correlations at long distance \txo{$\bar{P}_{d_{x^2\mathchar`-y^2}}$}.
We use both of \txo{the periodic-periodic (PP)
and antiperiodic-periodic (AP)
boundary conditions,
denoted by closed and open symbols, respectively}.
For the comparison, \txo{the data of the previous study on the signle-layer $t$-$t'$ Hubbard model~\cite{ido2018competition}
with $L=24$ and the AP boundary condition
are also shown by open circles.}}
\txcy{The thick light-red lines show linear interpolation of the superconducting correlations for $L=22$ with PP boundary condition.}}
        \label{fig:energy_bi_PP_U10}
\end{figure*}

\subsubsection{Many-body chemical potential}
\txc{Chemical potential for the $N$-electron interacting system is
evaluated by the following formula,
\begin{eqnarray}
    \mu({\delta=1-N/L^2})= \frac{E(N+\mathit{\Delta} N)-E(N-\mathit{\Delta} N)}{2\mathit{\Delta} N}, \label{eq:chemi_mu}
\end{eqnarray} 
where
$\mathit{\Delta} N$ is a positive integer \txo{much smaller} than $N$ ($\mathit{\Delta} N \ll N$).
Although it is ideal to set $\mathit{\Delta} N = 1$ and take the thermodynamic limit, $N\rightarrow +\infty$,
{$\mathit{\Delta} N$ is chosen to satisfy the closed shell condition} for the sake of the optimization
of the variational wave function.}

\subsection{\txo{Parameters for convergence}}
\subsubsection{\txo{System size}}
\txc{In the present mVMC simulation,
the number of sites per layer
is $\Ns=L\times L$ with $L=16, 18, 20$, and 22.
{We will use periodic-periodic (PP) and anti-periodic-periodic (AP) boundary conditions
in the following calculations.
In the PP boundary condition,
the periodic boundary condition is taken along both of the $x$ and $y$ directions.
On the other hand, in the AP boundary condition,
the anti-periodic boundary condition is taken along the $x$ and 
while the periodic bounary condition is taken along the $y$ directions.
}} 

\subsubsection{\txo{Monte Carlo samplings}}
\txc{In the variational Monte Carlo simulations,
the Markovian chain Monte Carlo sampling
is used to sample the real-space electron configuration $\ket{x}$,
where the probability that generates the Markovian chain is proportional to $\left| \left\langle x\right| \left. \psi\right\rangle \right|^2$.
By using the set of the sampled real-space configurations, $\Gamma_{\rm MC}$, we estimate the expectation value of an operator $\hat{O}$ as,}
\eqsa{
\frac{
\left\langle \psi\right| \hat{O} \left| \psi\right\rangle
}{
\left\langle \psi \right| \left. \psi\right\rangle
}
&=&
\sum_{x}
\frac{
\left\langle x\right| \hat{O} \left| \psi\right\rangle
}{
\left\langle x\right| \left. \psi\right\rangle
}
\frac{
\left| \left\langle x\right| \left. \psi\right\rangle \right|^2}
{
\left\langle \psi \right| \left. \psi\right\rangle
}
\non\\
&\simeq&
\frac{1}{N_{\rm MC}}
\sum_{x \in \Gamma_{\rm MC}}
\frac{
\left\langle x\right| \hat{O} \left| \psi\right\rangle
}{
\left\langle x\right| \left. \psi\right\rangle,
}
}
\txc{where $N_{\rm MC}$ is the number of the Monte Carlo steps or the number of the sampled real-space configurations.
In both of the optimization of the variational parameters and the evaluation of the observables,
we set $N_{\rm MC}=4\times 10^4 - 6.4 \times 10^4$.} 

\subsubsection{\txo{Variance extrapolation}}
\txo{To achieve the exact eigenvalues and physical quantities from variational approaches,
the variance extrapolation has been employed~\cite{PhysRevB.48.12037}.
However, it has been demonstrated that the
variance extrapolation does not significantly affect $P_{d}$~\cite{darmawan2018stripe}
in the Hubbard model.
Therefore, in the present paper, we do not perform the variance extrapolation to save the computational costs.}

\section{Results} \label{Sec:Results}
\begin{figure*}[htbp]
     \centering
    \includegraphics[width=0.8\textwidth]{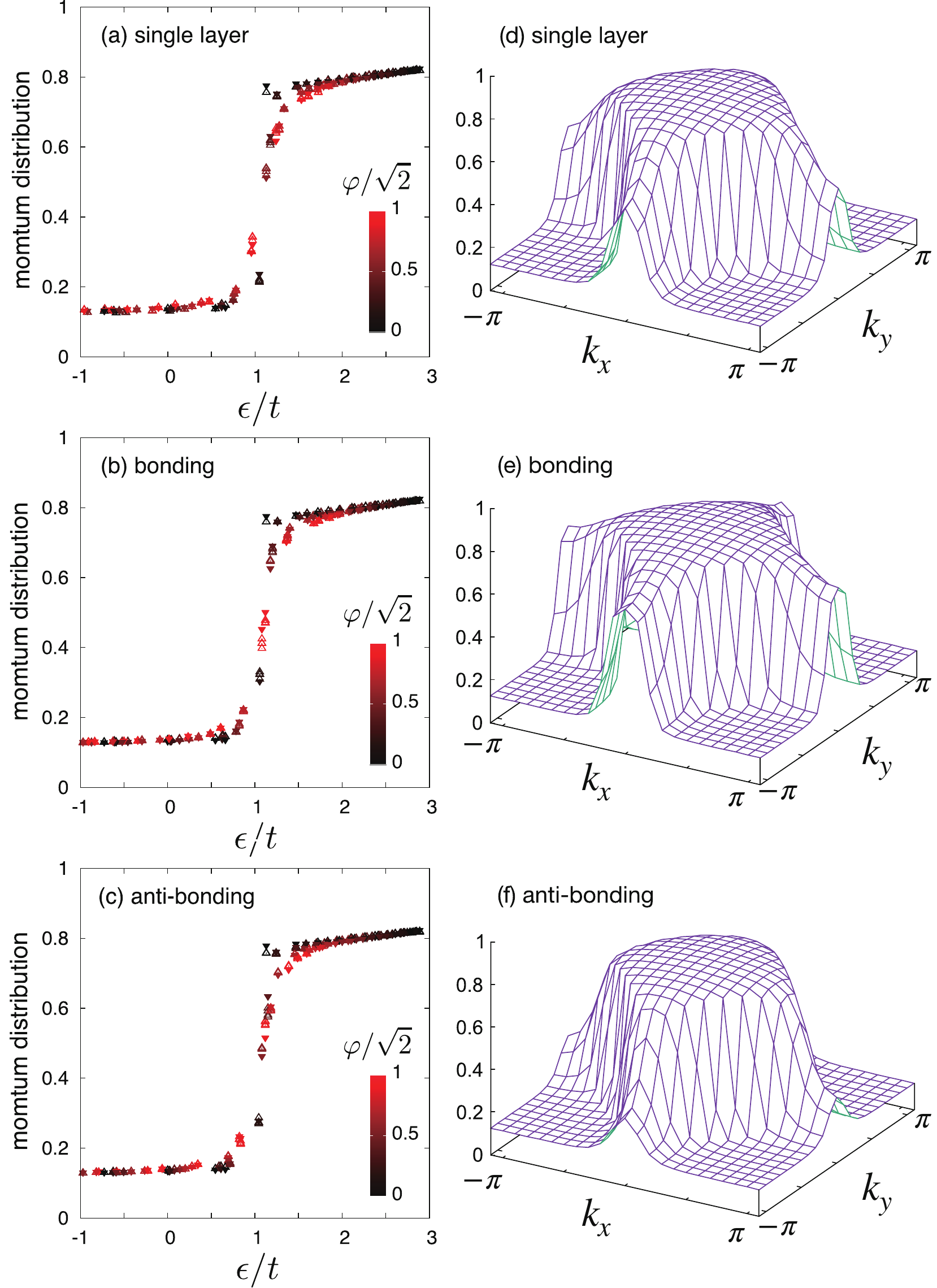}
     \caption{\txo{Momentum distribution and results of regression
by the fitting function $\widetilde{n}_{\bvec{k}}^{\pm}$ \txcy{[see Eq.~(\ref{eq:n_k_regression_model})]} at $\delta \simeq 0.22$.}
\txo{In comparison with the single layer results \txcy{for $t'/t = -100/360$} shown in (a), the results for the bonding and antibonding distribution,
$n_{\bvec{k}}^{+}$ and $n_{\bvec{k}}^{-}$, are shown in (b) and (c), respectively.
The momentum distributions obtained by the mVMC simulations are shown by upward open triangles while
the optimized fitting functions are shown by downward closed triangles.}
\txo{Here, $(k_x,k_y)$ dependence of $n_{\bvec{k}}^{\pm}$ is transformed into $(\epsilon, \varphi)$, where $\epsilon = \xi_{\bvec{k}}^{\pm}$ and $\varphi = \cos k_x - \cos k_y$}.
\gr{In the right \txcy{column,
the $\bm{k}$ dependences of the momentum distributions are shown for $L=22$.}}
\txcy{While the panel (d) shows the results for the single-layer system,
the panels (e) and (f) show the momentum distribution for the bonding and antibonding band of the bilayer system, respectively.}
}
     \label{fig:regression_n_of_k}
\end{figure*}

\begin{figure*}[htb]
     \centering
    \includegraphics[width=1.0\textwidth]{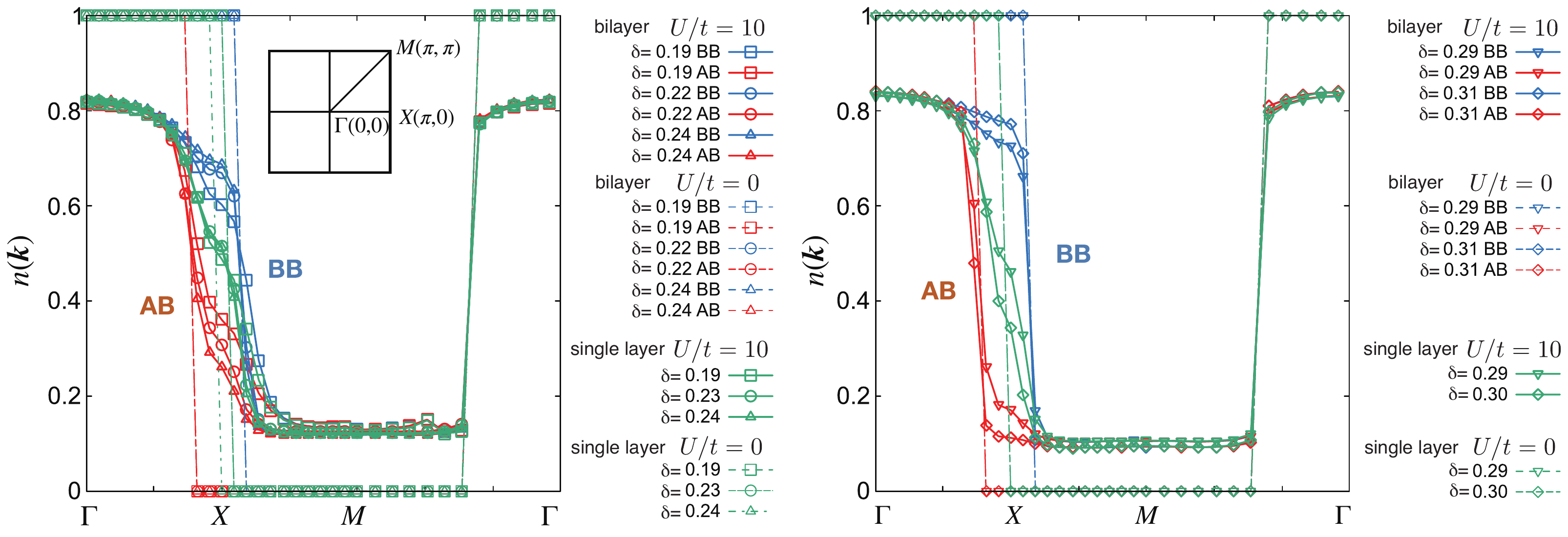}
     \caption{\txc{Momentum distribution \txcy{along symmetry lines} at
{$0.19\le\delta\le0.24$ (left panel) and $0.27\le\delta\le0.31$ (right panel)}. 
\gr{While the green open symbols with solid lines denote the momentum distribution of the single-layer $t$-$t'$ Hubbard hamiltonian for $t'/t=-100/360$, $U/t=10$, and $L=22$ with the PP boundary condition,
the red open symbols (blue open symbols) with solid lines
denote the results for the momentum distribution, $n_{\bm{k}}^{-}$ ($n_{\bm{k}}^{+}$), 
of the bilayer $t$-$t'$ Hubbard hamiltonian for $t'/t=-100/360$, $t_{\rm bi}/t=110/360$, 
$U/t=10$, and $L=22$ with the PP boundary condition.
\txcy{The open squares, circles, upward triangles, downward triangles, and diamonds denote the data of the bilayer system
at $\delta = 0.19$, $0.22$, $0.24$, $0.29$, and $0.31$, respectively. 
For the single-layer system, the  open squares, circles, upward triangles, downward triangles, and diamonds denote the data
at $\delta = 0.19$, $0.23$, $0.24$, $0.29$, and $0.30$, respectively.}
The momentum distribution of the non-interacting systems ($U/t=0$) is also plotted with dashed lines and open symbols.
\txcy{Here, we use the symbols to distinguish the doping levels in the same manner as for $U/t =10$.
The} $\bm{k}$ dependence of the momentum distribution is shown along the symmetry lines that connect
the symmetry points, $\Gamma$ $(0,0)$, $X$ $(\pi,0)$, and $M$ $(\pi,\pi)$.
}}}
     \label{fig:momentum_distribution}
\end{figure*}

\begin{figure}[htbp]
    \centering
    \includegraphics[width=0.4\textwidth]{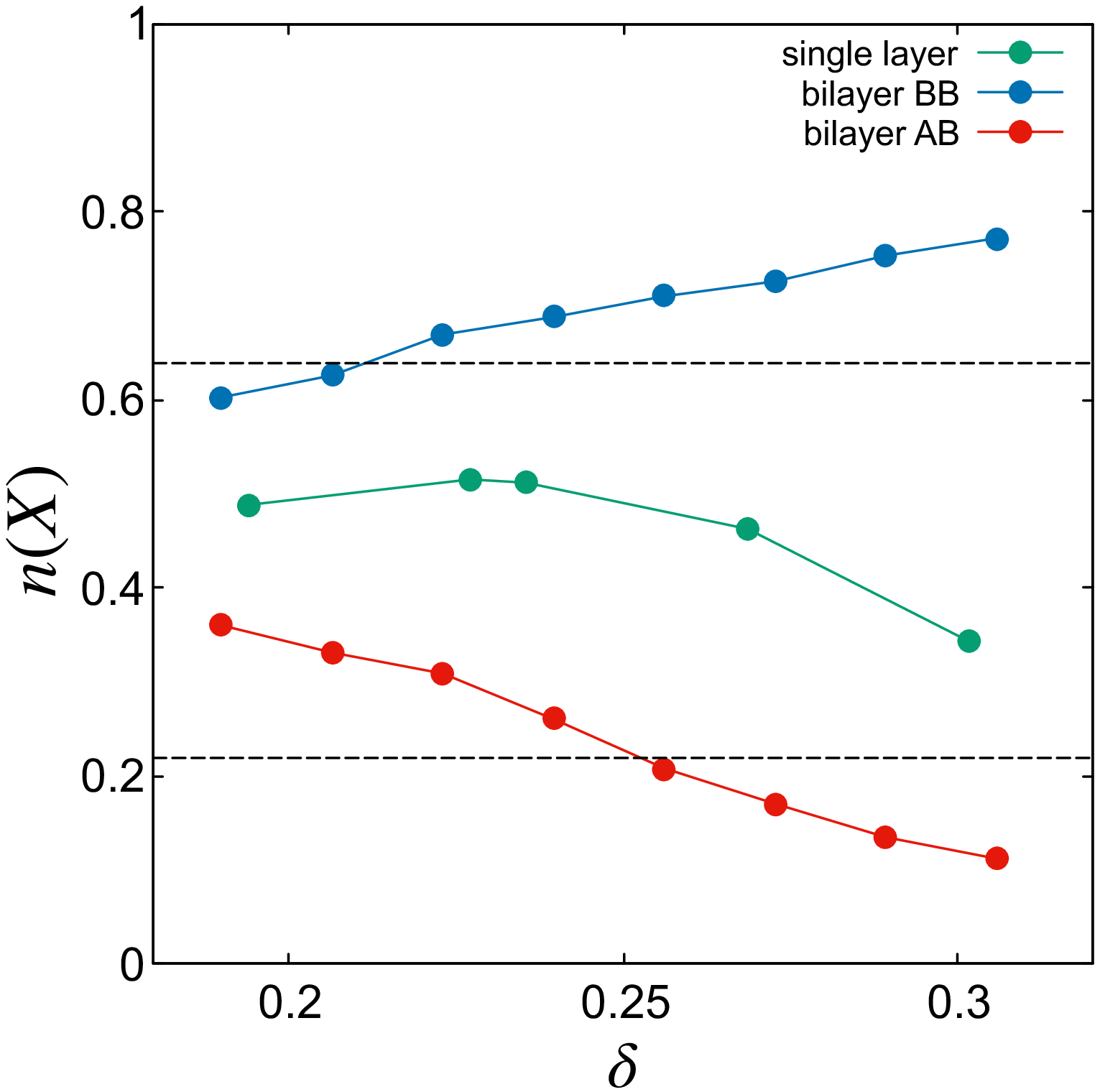}
    \caption{\gr{\txcy{Doping} dependence of the momentum distributions at \txcy{the} $X$ point for the single-layer and bilayer $t$-$t'$ Hubbard \txcy{hamiltonians}. 
\txcy{{The blue and red closed circles denote the doping dependence of the momentum distribution $n_X$ for the bonding and anti-bonding bands, respectively,
while the green closed circles denote $n_X$ for the single-layer system.}
    \txcy{The two dashed horizontal} lines show the lower limit $\sim 0.22$ and the upper limit $\sim 0.64$ \txcy{given by} the inequality \txcy{Eq.~(\ref{eq:inequality})}, respectively.} } }
    \label{fig:nk_X}
\end{figure}

\txc{While
the direct evaluation of 
the superconducting critical temperature $T_{\rm c}$
is beyond the scope of the present numerical algorithm,
there are observables that closely correlate with $T_{\rm c}$.}
\txo{The superconducting correlation function is a simple physical quantity
that correlate with $\Tc$ while it is hard to directly observe.} 
\txo{In contrast, the amplitude of the superconducting gap function is an observable
closely related to $\Tc$
while it is hard to simulate.}
\txc{From the ARPES measurements,
the optimal critical temperature \txo{$T_{\rm c}^{\rm opt}$}
correlates with the $d$-wave gap amplitude $\Delta_{0}$ around the nodal region~\cite{PhysRevLett.104.227001},
where $\Delta_{0}$ is \txo{determined} by fitting
\txo{the model function, $\Delta_0 \left(\cos k_x - \cos k_y \right)$, to} 
the experimentally observed SC gap
in the nodal region.}

\txo{In this section, first, we will introduce our results of
the doping dependence of the superconducting correlations
in comparison with the spin correlations.
From the momentum distribution, then, we extract the information of the superconducting gap function.}

\subsection{\txo{Spin and superconducting correlations}}

\txc{First, we examine the doping dependence of the superconductivity in
the bilayer $t$-$t'$ Hubbard hamiltonian, in comparison with
that of the single-layer counterpart.
As we discussed in Sec.~\ref{sec:single_CuO2},
we focus on the uniform \txo{superconducting phase and the antiferromagnetic} phase in the following.}

We numerically calculated \txc{the spin structure factor [Eq.~(\ref{eq:spin_cor})] and the intralayer SC correlation function
with the simple $d_{x^2\mathchar`-y^2}$ form factor [Eqs.~(\ref{eq:pair_cor}), (\ref{eq:Delta_s}), and (\ref{eq:form_factor})]}.
\gr{Since the inversion symmetry exists, physical quantities
\txcy{in the two layers
are same}
except for \txcy{those at} the certain dopings in the low-doping region \txcy{as} explained in Sec.~\ref{subsec:charge_fluc}.
\txcy{Therefore}, we omit \txcy{the layer indix} $\alpha$ of the physical \txcy{quantities below}.}
\txo{Typical $\bm{r}$ dependences of $P_{d_{x^2\mathchar`-y^2}}(\bm{r})$ are shown in Fig.~\ref{fig:superconducting_correlations}.
In the stable superconducting phase, the superconducting correlation converges to a constant at a long distance, $|\bm{r}| \gg 3a$.}
\txc{Figure~\ref{fig:energy_bi_PP_U10} shows the doping dependence of the SC correlations at long distance
[Eq.~(\ref{eq:pair_cor_ave})] and the peak values of the spin structure factor [Eq.~(\ref{eq:spin_cor})]
for
the bilayer
$t$-$t'$ Hubbard hamiltonian with $t'/t = -100/360$ and \txo{$U/t=10$} (see Table~\ref{table:hopping})
in comparison with those for the single-layer $t$-$t'$ Hubbard \txo{model} for $t'/t=-0.3$ \txo{and $U/t=10$~\cite{ido2018competition}}}.

\txc{As found in the literature on the single-layer
Hubbard hamiltonian~\cite{giamarchi1991phase,capone2006competition,aichhorn2007phase,yokoyama2012crossover,misawa2014origin,Zheng2016_PhysRevB.93.035126,ido2018competition},
the antiferromagnetic state, stabilized around the half-filling, becomes unstable upon increasing doping,
and the superconducting state becomes stable for the larger doping.
\txcy{In particular,
in the $t$-$t'$ Hubbard hamiltonian with $t'/t\sim -0.3$,}
the superconducting state is stable for $\delta \gtrsim 0.2 $ while 
the antiferromagnetic state in the low-doping region $\delta \lesssim 0.2$}~\cite{misawa2014origin,ido2018competition}.
It is common for both
\txo{single-layer and bilayer systems}
that the SC correlation \txc{develops upon} increasing $\delta$
and disappears after reaching a peak.

\txc{The superconducting correlations at long distance both in the single-layer and bilayer hamiltonians are
almost same at the optimal doping as shown in Fig.~\ref{fig:energy_bi_PP_U10}.
\txo{The} difference between the
single-layer and bilayer hamiltonians
becomes evident in the superconducting correlations at the larger doping region.
In the bilayer system,
$\overline{P}^{\alpha}_{d_{x^2\mathchar`-y^2}}$ 
is significantly small \txcy{in} 
the overdoped region {($\delta \gtrsim 0.25$)} in comparison with
\txo{$\overline{P}^{\alpha}_{d_{x^2\mathchar`-y^2}}$} 
in the single-layer system.}

\txc{The reduction at the overdoped region is attributed to the van Hove singularity of the band dispersion.}
\tby{In the single-layer $t$-$t'$ Hubbard hamiltonian,
the superconducting gap opens across the van Hove singularity in the normal state,
in the wide range of the hole doping as illustrated in the following section \ref{sec:van_Hove}.
In contrast, the superconducting gap does not involve the van Hove singularity in the bilayer $t$-$t'$ Hubbard hamiltonian
at the overdoped region.}

\subsection{Momentum distribution}

\txo{The momentum distribution $n_{\bm{k}}^{\pm}$
contains information about the single-particle spectrum.
The Fermi liquid theory shows that
$n_{\bm{k}}^{\pm}$ has discontinuity at the Fermi momentum $\bm{k}_{\rm F}^{\pm}$ in the metallic ground state.
The formation of the superconducting gap at the Fermi momentum removes the discontinuity.
However, even in the superconducting phase,
the momentum distribution functions shows the remnant of the discontinuous jump at the Fermi surface
and the information of the superconducting gap function.}

\subsubsection{Superconducting gap} \label{sec:SCgap}

\txo{\txc{While $n_{\bm{k}}^{\pm}$ shows discontinuity at $\bm{k}_{\rm F}^{\pm}$ in the metallic phase,}
\txcy{$n_{\bm{k}}^{\pm}$} is smoothened and the discontinuity disappears in the superconducting phase.
The amplitude of the gap function is reflected in
the smoothness of $n_{\bm{k}}^{\pm}$ {at the normal-state Fermi momentum $\bm{k}_{\rm F}^{\pm}$
that
satisfies $\left. \xi_{\bm{k}}^{\pm}\right|_{\bm{k}=\bm{k}_{\rm F}^{\pm}}=0$}.
The gradient of the momentum distribution $n_{\bm{k}}^{\pm}$ at $\bm{k}_{\rm F}^{\pm}$
contains information on the gap function $\Delta^{\pm} (\bm{k})$.
When the gap function $\Delta^{\pm} (\bm{k})$ becomes finite, the gradient of the momentum distribution
and the gap function has the following approximate relationship, 
$\left. 1/\|\nabla_{\bm{k}}n_{\bm{k}}^{\pm}\| \right|_{\bm{k}=\bm{k}_{\rm F}^{\pm}}\sim
\Delta^{\pm} (\bm{k}_{\rm F}^{\pm}) / v_{\bm{k}_{\rm F}^{\pm}}$}\txo{,
where $v_{\bm{k}_{\rm F}^{\pm}}$ is the Fermi velocity of the non-interacting band dispersion $\xi_{\bm{k}}^{\pm}$ at $\bm{k}_{\rm F}^{\pm}$}.

\txc{Instead of taking the derivative of the finite-size discrete data,
we perform a regression of $n_{\bm{k}}^{\pm}$ by
\txo{introducing a model function.}
\txo{The simplest model of the momentum distribution is given by the mean-field
ansatz with $d$ wave superconducting gap.
If the $\bm{k}$ dependence is captured by the mean-field ansatz,
the detailed $\bm{k}$ dependence of the momentum distribution will be simplified
by introducing a new set of the valuables, $(\epsilon, \varphi)$,
where $\epsilon = \epsilon_{\bm{k}}^{\pm}$ is
the single particle energy and $\varphi =\left|\cos k_x - \cos k_y\right|$ is
the angle-dependence of the $d$ wave superconducting gap function.}
\txo{Here, 
by taking into account the mean-field $(\epsilon,\varphi)$ dependence and the Fermi-liquid-like renormalization,
we introduce a phenomenological function $\widetilde{n}_{\bm{k}}^{\pm}$
defined
below.}}

\txcy{The phenomenological function is defined by combining a mean-field BCS momentum distribution
and smooth background as,}
\eqsa{
\widetilde{n}_{\bm{k}}^{\pm}&=&
\txo{n_{\rm b} \left(\epsilon_{\bm{k}}^{\pm},\mu_1,\tau_1 \right)}
\non\\
&&+
\zeta
n_{\rm MF}\left(\epsilon_{\bm{k}}^{\pm},\cos k_x - \cos k_y,\Delta_0,\mu_2\right),
\label{eq:n_k_regression_model}
}
where \txo{a smooth background is given by  
$n_{\rm b} (\epsilon,\mu_1,\tau_1) = n_0 
+ n_1 \exp [(\epsilon - \mu_1)/\tau_1]$,
and}
$\epsilon_{\bm{k}}^{\pm}$ is the non-interacting band dispersion.
Here, $n_0$, $n_1$, $\mu_1$, $\tau_1$, $\Delta_0$, and $\mu_2$ are fitting parameters. 
\txcy{The exponential function in the smooth background is introduced
to reproduce positive-definite and non-linear $\epsilon/t$ dependence beyond the following mean-field part.}
The mean-field momentum distribution function, $n_{\rm MF}$, is given by,
\eqsa{
n_{\rm MF}\left(\epsilon,\varphi,\Delta,\mu\right)
=
\frac{\sqrt{\left(
\Delta \varphi
\right)^2 + 
\left(\epsilon - \mu\right)^2 } 
- \epsilon + \mu }{2\sqrt{
\left(
\Delta \varphi
\right)^2 + \left(\epsilon - \mu\right)^2 } },
}
which follows the momentum distribution function of the BCS mean-field wave function,
\eqsa{
\sum_{\sigma}
\frac{
\bra{\phi_{\rm SC}}
{c_{\bm{k}\sigma}^{\pm}}^{\dagger}
c_{\bm{k}\sigma}^{\pm}
\ket{\phi_{\rm SC}}
}
{
\left\langle \phi_{\rm SC} \right|\left. \phi_{\rm SC} \right\rangle
}
&=&
2\left(v_{\bm{k}}^{\pm}\right)^2
\non\\
&=&
\frac{
\sqrt{\left(\xi_{\bm{k}}^{\pm}\right)^2+\left[\Delta_{\mathrm{SC}}^{\pm}(\bm{k})\right]^2}
-\xi_{\bm{k}}^{\pm}}{
\sqrt{\left(\xi_{\bm{k}}^{\pm}\right)^2+\left[\Delta_{\mathrm{SC}}^{\pm}(\bm{k})\right]^2}
}
.\non\\
}
\txo{Here, the coefficient $v_{\bm{k}}^{\pm}$ is given in Eq.~(\ref{eq:v_BCS}).}

\txo{We fit the function $\widetilde{n}_{\bm{k}}^{\pm}$ to
the numerical data of $n_{\bm{k}}^{\pm}$ \txcy{by least squares}
at a doping $\delta \simeq 0.22$,
where both the single-layer and bilayer $t$-$t'$ Hubbard hamiltonians
show stable superconductivity as shown in Fig.~\ref{fig:regression_n_of_k}}.
{To utilize data at dense momentum points,
here, we use the numerical data for three different system sizes,
$L=18, 20$, and $22$, simultaneously, to find a single fitting function.
We also perform a similar fitting for the data from the single-layer Hubbard hamiltonian.
For each system size $L$, we choose the electron number $N_{\rm e}$
to make the doping $\delta = 1 - N_{\rm e}/L^2$ close to $0.22$,
which is summarized in Table~\ref{table_number_electron}.
\txo{Here, we only use the data for $\varphi < \sqrt{2}$ to focus on the nodal region.}
\txcy{To quantify the performance of the regression, we estimate the root mean square errors of
the fitting functions $\widetilde{n}_{\bm{k}}^{\pm}$
within a range, $0.5 < \epsilon/t < 1.5$.
While the root mean square error is 0.01 for the single-layer system,
the errors are 0.02 for both the bonding and anti-bonding bands of the bilayer system.
Therefore, the regression is reasonable.}}



{By the regression, we extract the gap function around the nodal region
as shown in Table~\ref{table:Delta_Pd}.
\txcy{Here, we estimate the errors in the fitting parameters by the bootstrap samples~\cite{efron1992bootstrap}.}
\txo{The amplitude of the gap function, $\Delta_0$, for the bilayer system is quantitatively similar to that for the single-layer system.
The amplitude of the gap functions in the BB and AB bands is also indistinguishable.}
{The recent ARPES measurement shows the SC gap in
the BB and AB bands are distinct \txo{around the antinodal region}~\cite{ai2019distinct}
while the previous measurement~\cite{feng2001bilayer}
could not distinguish
these SC gaps
in the overdoped Bi2212.
\txo{However, around the nodal region, the amplitude of the gap functions is almost identical \txcy{even in the recent measurement}, and, thus,}
\txcy{we conclude that} the recent ARPES observation is consistent with $\Delta_0$
obtained for the BB and AB.}}

{Then, we can estimate the effective attractive interaction through the following formula,
\eqsa{
V_d = 2\Delta_0/\sqrt{\overline{P}_{d_{x^2\mathchar`-y^2}}^{\alpha}}.
}
\txo{The effective interaction $V_d$ of the bilayer system is also similar to that of the single-layer system
as shown in Table~\ref{table:Delta_Pd}.}
\txo{These results are} consistent with the effective interaction,
$V_d=1.7t$,
estimated from the spectral weight of the Hubbard model at $\delta = 0.125$ and $U/t=8$
~\cite{PhysRevX.10.041023}.}

\begin{table}[htb]
\begin{center}
\caption{\txc{System size $L$, number of electrons $N_{\rm e}$, and doping $\delta$
used for estimating superconducting gap from momentum distribution functions}.
\label{table:Delta_Pd}}
\begin{tabular}{cccccccc}
&&&&&&&    \\
\hline
\hline
& & & \multicolumn{5}{c}{$L$} \\
\cmidrule{4-8}
& & & $18$ & & $20$ & & $22$ \\
\hline
single-layer & $N_{\rm e}$ & & $254$ & & $314$ & & $374$ \\
 & $\delta$ & & $0.216$ & & $0.215$ & & $0.227$ \\
\hline
bilayer & $N_{\rm e}$ & & $504$ & & $624$ & & $752$ \\
 & $\delta$ & & $0.222$ & & $0.22$ & & $0.223$ \\
\hline
\hline
\end{tabular}
\label{table_number_electron}
\end{center}
\end{table}

\begin{table}[htb]
\begin{center}
\caption{\txc{Superconducting gap estimated from momentum distribution functions
for $L=18, 20$, and 22, and
superconducting correlation at long distance for {$L=22$}}.
{We perform the regression with the AB band for the gap function $\Delta_0$ of
the bilayer system.}
\txcy{The errors in $\Delta_0$ and, thus, in $V_d$ are estimated by the regressions
for 100 of the bootstrap samples~\cite{efron1992bootstrap}.
The results of the regression with the original data are shown in brakets.}
\label{table:Delta_Pd}}
\begin{tabular}{cccc}
    &&&\\
\hline
\hline
& $\Delta_0 /t$ & $\overline{P}_{d_{x^2\mathchar`-y^2}}^{\alpha}$ & $V_d/t$ \\
\hline
single-layer & $0.20\pm 0.01$  & $0.0558$ & $1.7\pm 0.1$ \\ 
& $(0.193)$ && $(1.64)$\\
bilayer (AB) & $0.19\pm 0.01$ & $0.0526$ & $1.63\pm 0.09$  \\
& $(0.184)$ && $(1.61)$ \\
\txo{bilayer (BB)} & $0.19\pm 0.01$  & $0.0526$ & $1.66\pm 0.09$ \\
& $(0.188)$ && $(1.64)$ \\
\hline
\hline
\end{tabular}
\label{table_gap}
\end{center}
\end{table}

\subsubsection{Van Hove singularity}
\label{sec:van_Hove}

\txo{Even in the superconducting phase,
anomalous doping dependences of the physical quantities have often been attributed to
the Lifshitz transition, namely, changes in the Fermi-surface topology of the metallic phase.
When the Fermi surface shrink across the saddle points of the non-interacting band dispersion $\epsilon_{\bm{k}}^{\pm}$,
which are located at the $X$ points [$(\pi/a,0)$ and $(0,\pi/a)$],
upon hole doping, the Lifshitz transition occurs.
The anomalies around the Lifshitz transition originate from the van Hove singularity of the density of states.
The stability of the superconductivity may be affected by the van Hove singularity, even in the strongly correlated electron systems.}

\txo{By exploiting the model function $\widetilde{n}_{\bm{k}}^{\pm}$ [Eq.~(\ref{eq:n_k_regression_model})],
we will analyze the impacts of the van Hove singularity on the superconductivity.}
\txo{\txcy{In particular}, we examine whether the formation of the superconducting gap involves the van Hove singularity.}
\txo{An inequality,
\eqsa{
\left|\epsilon_{\bm{k}}^{\pm} -\mu_2 \right| \lesssim \left|\Delta_0 (\cos k_x - \cos k_y ) \right|\label{eq:inequality}
}
offers a simple criterion for determining whether the single-particle spectrum at $\bm{k}$ is involved in
the formation of the superconducting gap.
From the momentum distribution, we can determine whether the inequality Eq.~(\ref{eq:inequality}) holds at a given momentum.
Thus, we can determine whether the saddle point $X$ is involved in the gap formation or not.
At least, the inequality is easily transformed into a condition on the mean-field component $n_{\rm MF}$, which is determined by $\left|\epsilon_{\bm{k}}^{\pm} -\mu_2 \right|$
and $\left|\Delta_0 (\cos k_x - \cos k_y ) \right|$, as
\eqsa{
\frac{\sqrt{2}-1}{2\sqrt{2}}
\lesssim
n_{\rm MF} \lesssim
\frac{\sqrt{2}+1}{2\sqrt{2}}.
}
Then, if the inequality Eq.~(\ref{eq:inequality}) holds at $\bm{k}$, the momentum distribution satisfies
\eqsa{
n_{\rm b} + \frac{\sqrt{2}-1}{2\sqrt{2}}\zeta \lesssim n_{\bm{k}} \lesssim n_{\rm b} + \frac{\sqrt{2}+1}{2\sqrt{2}}\zeta.
\label{eq:condition_on_nk}}
}

\txo{To utilize the condition Eq.~(\ref{eq:condition_on_nk}),
we need to determine the smooth back ground $n_{\rm b}$ and the renormalization constant $\zeta$.
Here, we assume that $n_{\rm b}$ and $\zeta$ around the Fermi momentum weakly depend on $\varphi$.
Then, we can estimate $n_{\rm b}$ and $\zeta$ from the momentum distribution along the nodal line, $k_x = k_y$,
or $\varphi = 0$.
%
From data with $\varphi=0$ in Fig.~\ref{fig:regression_n_of_k},
the momentum distribution shows discontinous jump from $n_{\bm{k}_{\rm F}^{\pm}} = n_{\rm b} \sim 0.13$
to $n_{\bm{k}_{\rm F}^{\pm}} = n_{\rm b} + \zeta \sim 0.73$ at $\delta \sim 0.22$ and $U/t=10$,
in both the single-layer and bilayer systems.
Therefore, the ineqauality Eq.~(\ref{eq:inequality}) holds,
the momentum distribution satisfies $0.22 \lesssim n_{\bm{k}} \lesssim 0.64$.}

\txo{From the doping dependece of the momentum distribution,
here, we determine whether the van Hove singularity is involved in the gap formation at each doping.
As shown in Fig.~\ref{fig:momentum_distribution}, the momentum distribution at the saddle point $X$,
$n_{X}$, for the single-layer system remains in the range of 0.3 to 0.6 and clearly satisfies the condition, $0.22 \lesssim n_{\bm{k}} \lesssim 0.64$
for $\delta \lesssim 0.3$.
In contrast, the momentum distribution of the bonding band $n_{\bm{k}}^{+}$ increases beyond $0.64$ upon increasing hole doping
while $n_{\bm{k}}^{-}$ decreases below $0.22$ as shown in Fig.~\ref{fig:nk_X}.}

\txo{Therefore,
we conclude that, in the bilayer $t$-$t'$ Hubbard hamiltonian,
the superconducting gap involves the van Hove singularity only within a small range of the doping
and does not involve the van Hove singularity at the overdoped region, $\delta \gtrsim 0.25$,
while, in the single-layer $t$-$t'$ Hubbard hamiltonian,
the superconducting gap opens across the van Hove singularity
in the wide range of the hole doping, $0.19 \lesssim \delta \lesssim 0.3$.
}
{Due to the bilayer band splitting, the Fermi surfaces of BB and AB cannot simultaneously exploit the high density of
\txo{states around the $X$ point}.}
{This inhibits the formation of the superconducting gap for the overdoped region in the bilayer \txo{hamiltonian}.}
\txo{The importance of the van Hove singularity confirmed by the present simulation
seems to support the van Hove scenario~\cite{PhysRevLett.56.2732,PhysRevB.52.13611,MARKIEWICZ19971179}.}

\begin{figure*}[htb]
     \centering
    \includegraphics[width=1.0\textwidth]{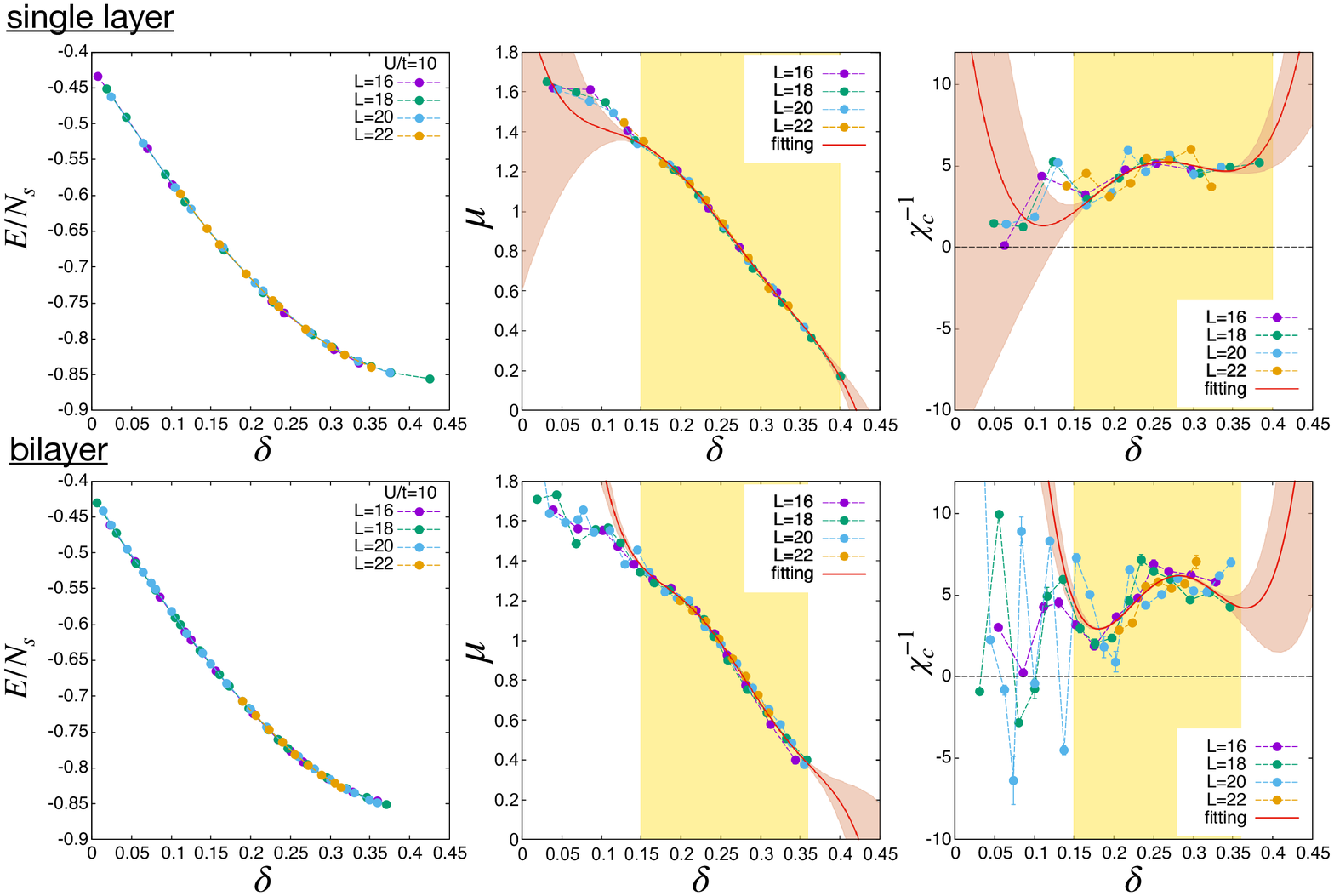}
     \caption{
\txc{Doping dependence of ground-state energy, chemical potential, and inverse of charge susceptibility \txcy{in units of $t$}.
While the upper panels show the results of the single-layer $t$-$t'$ Hubbard hamiltonian for $t'/t=-100/360$ and $U/t=10$,
the lower panels show the results for the bilayer $t$-$t'$ Hubbard hamiltonian for $t'/t=-100/360$, $t_{\rm bi}/t=110/360$,
and $U/t=10$.
The $\delta$ dependence of the ground-state energy per site, $E/N_{\rm s}$, is shown in the left panels.
The middle panels show the $\delta$ dependence of the chemical potential, $\mu (\delta)$, estimated by Eq.~(\ref{eq:chemi_mu}).
The derivative of $\mu (\delta)$,
which is the inverse of the uniform charge susceptibility, $\chi_{\rm c}^{-1}$, is plotted in the right panels.
In each panels, we plotted the data for $L=16$ (purple closed circles), $18$ (green closed circles), $20$ (cyan closed circles),
and $22$ (orange closed circles).}
\gr{The chemical potential $\mu (\delta)$ is fitted with \txcy{a} fifth-order polynomial for $0.15 < \delta$,
which is plotted as the red curve in the \txcy{top and bottom middle} panels.
The \txcy{derivatives} of the fitted lines \txcy{are} also plotted in the \txcy{top and bottom} right panels by the \txcy{red curves}.
The shaded red \txcy{belts} in the middle and right panels show the \txcy{standard deviation} (1$\sigma$) of the fitting \txcy{functions of $\mu$ and its derivative}
\txcy{estimated} by the \txcy{jackknife sampling~\cite{efron1992bootstrap}}.
The shaded yellow \txcy{regions} ($0.15<\delta\lesssim 0.4$) \txcy{indicate} the doping range \txcy{used} for the fitting \txcy{of} $\mu(\delta)$.} 
}
     \label{fig:E_chemical_chi_c}
\end{figure*}

\subsection{\tby{Charge fluctuations}}\label{subsec:charge_fluc}

\txc{The uniform charge susceptibility, $\chi_{\rm c}$, correlates with instability towards the superconductivity in the Hubbard model~\cite{misawa2014origin},
which is obtained by the derivative of the doping dependence of the chemical potential, $\mu (\delta)$,
as
\begin{eqnarray}
\txcy{\chi_c =-\left[d \mu (\delta) /d \delta\right]^{-1}.}
\end{eqnarray}
In Fig.~\ref{fig:E_chemical_chi_c},
the charge susceptibility is shown for the single-layer and bilayer $t$-$t'$ Hubbard hamiltonians,
which is obtained from the doping dependence of the ground-state energy and chemical potential.}
\txcy{To focus on the charge fluctuations around the superconducting phase,
we estimate $\chi_{\rm c}$
at the hole doping range $0.15\lesssim \delta \lesssim 0.35$,
where $\bar{P}_{d_{x^2\mathchar`-y^2}}$ is of the order of or larger than 0.01}.

\txcy{As evident in the middle and right panels in Fig.~\ref{fig:E_chemical_chi_c},
the numerical derivatives of the ground-state energy $E$ become noisy.}
\gr{\txcy{In particular, in the bilayer system} for $\delta \lesssim 0.15$,
$\chi_{\rm c}$ shows significant size dependence.
\txcy{We found that there} is a tendency of the interlayer polarization of charge/spin \txcy{at} the low-doping region for the bilayer $t$-$t'$ Hubbard model.
\txcy{In addition,}
the overlap matrix $S$ in the SR method~\cite{tahara2008variational} tends to
\txcy{be of low-rank
for $\delta \lesssim 0.15$\txcy{, which} inhibits the optimization of the \txcy{ground-state} energy
and exaggerates the errors in the numerical derivatives.
Therefore,
the inverse charge susceptibility $\chi_{\rm c}^{-1}$ given by the numerical derivative is
not reliable for the low doping region, $\delta \lesssim 0.15$.}}

\txcy{Even though there are the errors in $\chi_{\rm c}^{-1}$ for $\delta \gtrsim 0.15$
and the uniform charge fluctuations seems to be enhanced,
$\mu$ is a monotonically decreasing function of $\delta$.
Thus, there is no clear indication of the phase separation\txcy{,
while the clear tendency towards the phase separation was found} in 
the standard single-layer
Hubbard hamiltonian $(t'/t=0)$~\cite{misawa2014origin}.}

\txo{The uniform charge fluctuations
show the similar doping dependence and amplitude
in both the single-layer and bilayer $t$-$t'$ Hubbard hamiltonians.
Although $\chi_{\rm c}$ correlates with the instability towards
the superconductivity~\cite{misawa2014origin,Misawa2014_LaFeAsO},
$\chi_{\rm c}$ alone hardly explains the distinct doping dependence of $\bar{P}_{d_{x^2\mathchar`-y^2}}$
in the single-layer and bilayer systems.}

\txcy{Here, we note that $\chi_{\rm c}^{-1}$ obtained in the present study is smaller than the
weak-coupling random phase approximation result, $\chi_{\rm c}^{-1} = ( 2 \Pi_0 )^{-1} + U/2$ $(> U/2 = 5t)$,
where $\Pi_0$ is the bare polarization function (equal to the density of state per spin).
The substantial reduction of $\chi_{\rm c}^{-1}$ clearly exhibits
relevance of non-perturbative correlations such as local-field corrections.}

\section{Summary and discussion}
\label{Sec:Summary_and_Discussion}

\txc{In the present paper,
one of the simplest hamiltonians for the bilayer cuprates is studied
in comparison with the single-layer system.
Our numerical results on the superconducting correlations and gap functions of
the bilayer $t$-$t'$ Hubbard hamiltonian revealed that
{the adjacent Hubbard layer does not
make the superconductivity more stable},
which is in contrast to the higher $\Tc^{\rm opt}$ and larger $\Delta_0$ in Bi2212
than those in Bi2201~\cite{PhysRevLett.89.067005,damascelli2003angle}.}

\txo{Due to the bilayer splitting, 
it is hard to generate the superconducting gap across the van Hove singularity for
$\delta \gtrsim 0.25$, and, thus,
the superconducting correlation $P_{d_{x^2\mathchar`-y^2}}$ in the bilayer $t$-$t'$ Hubbard hamiltonian is
smaller than that in the single-layer system.
Since the relationship between $\Tc$ and $P_{d_{x^2\mathchar`-y^2}}$ is not so clear,
the superconducting gap around the nodal region, which correlates with $\Tc$, was directly examined in the present paper.
We analyzed the momentum distribution and extracted the gap amplitude by performing a regression.
When the amplitude of the superconducting gap is smaller than the finite-size gap in the energy spectrum,
it is impossible to extract the information of the gap function from the momentum distribution.
Therefore, to make our regression reliable, we focused on the doping at which $P_{d_{x^2\mathchar`-y^2}}$ is optimal and the gap amplitude
is expected to be maximum.
The gap amplitude $\Delta_0$ and effective attractive interaction $V_d$
were found to be similar in both single-layer and bilayer systems.
Therefore, we concluded that the adjacent Hubbard layer does not enhance the stability of the superconductivity.}

\txc{\txo{The present results show that there are relevant factors to the high critical temperatures of the multi-layer cuprates that are not taken into accont in
the bilayer $t$-$t'$ Hubbard hamiltonians.}}
{Remaining factors relevant to the stability of the superconductivity
would be \txo{the long-range Coulomb repulsion,
differences} between the Hubbard and CuO$_2$ layers, and
effects of impurities or dopants.}

{The long-range Coulomb repulsion
in {\it ab initio} hamiltonians
is relevant to the stability of the superconductivity~\cite{ohgoe2020ab},
as \txo{reviewed} in Sec.~\ref{sec:single_CuO2}.
In modern technologies for derivation of the {\it ab initio} hamiltonians,
the Coulomb repulsion in the low-energy degrees of freedom is estimated by
the constrained random phase approximation (cRPA)~\cite{aryasetiawan2004frequency}.
The adjacent CuO$_2$ layer introduces additional channels in
the cRPA or c$GW$ screening process of the long-range Coulomb repulsion. 
To examine the impacts of these screening channels on the superconductivity,
it is desirable to study {\it ab initio} hamiltonians of typical examples of
single-layer and bilayer cuprates, such as
Bi2201 and Bi2212, respectively.} 

As examined in the literature~\cite{PhysRevLett.105.057003,hirayama2018ab},
{there are cuprates in which physics inside a single CuO$_2$ layer \txo{is not} captured by the single-orbital Hubbard-type
hamiltonian.
{\it Ab initio} studies based on the constrained $GW$ (c$GW$) approximation~\cite{hirayama2013derivation,hirayama2017low}
revealed that
\txo{an} {\it ab initio} single-orbital hamiltonian partly failed to reproduce properties of (La,Sr)$_2$CuO$_4$
while it succeeded in reproducing those of Hg1201 (HgBa$_2$CuO$_{4+y}$)~\cite{hirayama2018ab,ohgoe2020ab}. 
Even though the single-layer Hg1201 is successfully described by the {\it ab initio} single-orbital hamiltonian,
the changes in the number of the adjacent CuO$_2$ layers may require multi-orbital hamiltonians,
such as $dp$ hamiltonians~\cite{PhysRevB.45.10032} or two orbital hamiltonians with $d_{z^2}$ orbitals~\cite{PhysRevLett.105.057003,PhysRevResearch.3.033157}.}


\txc{Explicit oxygen degrees of freedom, which is absent in the Hubbard-type hamiltonians,
will play an important role in the {\it ab initio} multi-orbital hamiltonians.
\txcy{In the literature on the material dependence of the superconducting critical temperature of the cuprates,
relevance of the apical oxygen to $T_{\rm c}$
has been studied~\cite{PhysRevB.43.2968} in particular,} 
which has been clarified further by a modern regression scheme~\cite{doi:10.1021/acs.jpclett.1c01442}}.

\txo{It is highly desirable to derive and analyze {\it ab initio} effective hamiltonians of the series of the typical multilayer cuprates, Bi$22(n-1)n$,
that take into account the $n$ dependence of the screening
and multi-orbital nature including the $p$ orbitals of the oxygen ions.
\txcy{The} difference between the infinite layer high-$\Tc$ cuprates~\cite{Smith1991},
\txcy{particularly}, (Sr$_{1-x}$Ca$_x$)$_{1-y}$CuO$_2$~\cite{Azuma1992},
and Bi$22(n-1)n$ of finite $n$ is also crucial to
identify the impact of the apical oxygen atoms on the stability of the superconductivity.}

{The disorder effect of impurities and dopants from the charge reservoir block next to the 
CuO$_2$ layer is detrimental to $\Tcmax$ \cite{PhysRevB.69.064512,uchida2014high}. 
Such disorder effect is weaker in the bilayer cuprates, because the disordered charge reservoir block exists only in one side of a CuO$_2$ layer.
It is further weakened in the $n$-layer cuprates ($n\ge 3$) more than single-layer cuprates, because inner CuO$_2$ layers are protected by outer CuO$_2$ layers from the disordered charge reservoir block.  For larger $n$-layer system, the inner CuO$_2$ surface is cleaner and higher $\Tcmax$ is realized \cite{doi:10.1143/JPSJ.81.011008}. 
In the Hubbard-like hamiltonian, the ideal clean situation is realized. 
The examination of the disorder effect in the multilayer hamiltonians in comparison with the single-layer hamiltonian is \txo{also} desirable in the future.}

{Difference between the present results and properties of a typical bilayer cuprate Bi2212
is not only in the stability of superconductivity but also in the stability of the antiferromagnetic phase
at the underdoped region.}
{In comparison with the experimental phase diagram shown in
Fig. 3 of Ref.~\citen{drozdov2018phase},
the critical doping at the overdoped limit is similar
while antiferromagnetic state, which competes with the superconducting state, becomes stable in
the bilayer $t$-$t'$ Hubbard hamiltonian
for the optimal and underdoped region.}


{Recently,
the authors of Ref.~\citen{kunisada2020observation} proposed that
the interlayer hoppings are irrelevant and each CuO$_2$ layers are independent,
based on their ARPES spectra of a five-layer cuprate Ba$_2$Ca$_4$Cu$_5$O$_{10}$(F,O)$_2$.}
{It seemingly contradicts the clear band splitting observed in the bilayer cuprates.
The momentum distribution functions of
the bilayer $t$-$t'$ Hubbard hamiltonian, 
$n_{\bm{k}}^{\pm}$, also clearly show the band splitting.
\txo{The observations of the band splittings in the bilayer ($n=2$) and five-layer ($n=5$)
seemingly contradict.
It is left for future studies to elucidate the origin of the contradiction by performing
simulations for $n\geq 3$.}}

\begin{acknowledgment}
{\bf Acknowledment}\quad We thank Yukitoshi Motome for stimulating discussion.
Y. Y. thanks Masatoshi Imada for critical and helpful comments
\txcy{and Takahiro Misawa for insightful discussion}.
This research was supportd by MEXT as
``Basic Science for Emergence and Functionality in Quantum Matter
- Innovative Strongly-Correlated Electron Science by Integration of Fugaku and Frontier Experiments -" (JPMXP1020200104) as a program for promoting researches on the supercomputer Fugaku, 
supported by RIKEN-Center for Computational Science (R-CCS) through HPCI System Research Project (Project ID: hp210163 and hp220166).
The computation in this work has been done using the computational resources of
the supercomputer Fugaku provided by the R-CCS and
the facilities of the Supercomputer Center, the Institute for Solid State Physics, the University of Tokyo.
\txcy{A. I. was financially supported by Quantum Science and Technology Fellowship Program (Q-STEP).
Y. Y. was supported by JSPS KAKENHI (Grant No. 20H01850).}
\end{acknowledgment}

\setcounter{equation}{0} 
\renewcommand{\theequation}{\Alph{section}·\arabic{equation}}
\setcounter{figure}{0} 
\renewcommand{\thefigure}{\Alph{section}·\arabic{figure}}
\setcounter{table}{0} 
\renewcommand{\thetable}{\Alph{section}·\arabic{table}}

\appendix
\label{chap:appendix}

\section{Benchmarking}
\label{chap:benchmark}
\begin{figure}[htbp]
    \centering    
    \includegraphics[width=0.4\textwidth]{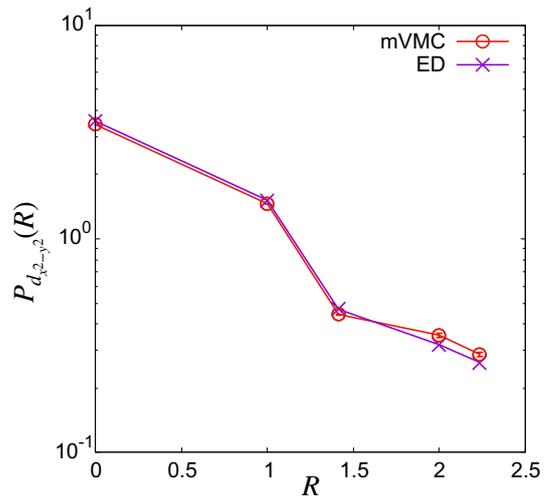}
    \caption{\gr{Comparison between mVMC and exact diagonalization (ED) superconducting correlation in $\Ns=4\times2$ bilayer Hubbard model at half-filling \txcy{with} $U/t=4$.} }
    \label{fig:Pd_EDvsVMC}
\end{figure}

\txo{To examine an accuracy of the present mVMC wave function Eq.~(\ref{eq:wf}),
we compare the ground-state energy, the peak value of \txcy{spin structure factor $S(\bvec{Q})$} and
the superconducting correlation $P_{d_{x^2\mathchar`-y^2}}(R)$ obtained by mVMC with \txcy{those} \txo{by} the exact diagonalization
\gr{for the bilayer Hubbard model defined \txcy{in} Eq.~(\ref{eq:bilayerHubbard}) with the same hopping matrices as Table \ref{table:hopping}.}
\gr{In Table~\ref{table_benchmark} and Fig.~\ref{fig:Pd_EDvsVMC}, we show the results of mVMC and the exact diagonalization at the half-filling ($\delta=0$) for $U/t=4$.}
The number of sites used in the comparison is $\Ns=4\times 2$ per layer and the bounary condition is periodic-periodic (PP).
\txcy{We performed the exact diagonalization by using an open source software for quantum lattice models, $\mathcal{H}\Phi$~\cite{kawamura2017quantum}.}
\txcy{The relative errors in $E/t$ and the peak value of $S(\bvec{Q})$ in the present mVMC calculations are 2\%.}}
\txcy{Here, the error in the ground-state energy $E/t$ is larger than that for the single-layer Hubbard model~\cite{misawa2014origin}.
The larger error is attributed to the energy gain by the spin quantum-number projection $\mathcal{L}^{S}$
since $\mathcal{L}^{S}$ is omitted for the bilayer system to save the computational cost.
Even when $\mathcal{L}^{S}$ is employed, the energy gain by the spin \txcy{quantum-number} projection in the bilayer system is smaller than that \txcy{in the single-layer system}.}
\begin{table}[htb]
\begin{center}
\caption{Ground-state energy $E/t$ and physical quantities obtained by mVMC and exact diagonalization (ED).}
\begin{tabular}{ccc}
    &&\\
\hline
\hline
$N_{\rm s}=4\times 2$& {\rm mVMC} & {\rm ED} \\
\hline
$E/t$ & $-20.27 \pm 0.01$ & -20.657 \\
$S(\bm{Q})/\Ns$ & $0.05902 \pm 0.00001$ & 0.05818\\
\hline
\hline
\end{tabular}
\label{table_benchmark}
\end{center}
\end{table}

\section{\gr{Effect of the spin quantum-number projection on the superconductivity}} \label{chap:SPvsNP}
\begin{figure}[htbp]
    \centering
    \includegraphics[width=0.4\textwidth]{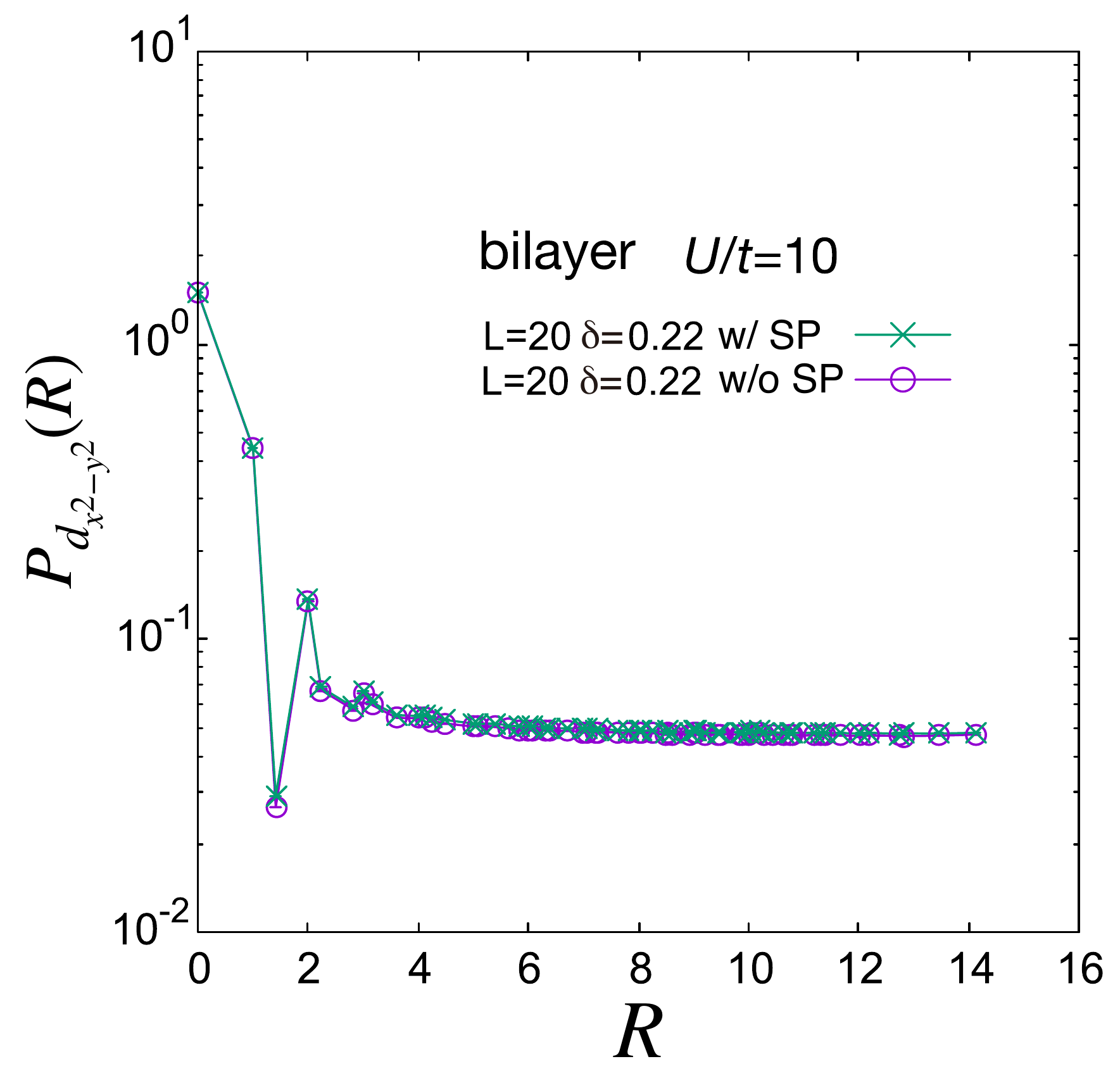}
    \caption{\txcy{Comparison of the \txcy{superconducting} correlation function $P_{d_{x^2\mathchar`-y^2}}(\bm{r})$ with the spin quantum-number projection $\mathcal{L}^{S}$ (w/ SP) and without $\mathcal{L}^{S}$ (w/o SP) for the bilayer $t$-$t'$ Hubbard model at the hole doping $\delta=0.22$ and $U/t=10$.
The cross symbols denote the results with $\mathcal{L}^{S}$ and the open circles denote
the results without $\mathcal{L}^{S}$}.}
    \label{fig:Pd_SPvsNP}
\end{figure}

\gr{To examine the effect of spin quantum-number projection $\mathcal{L}^{S}$ on the superconductivity,
we compare the intralayer superconducting correlation functions $P_{d_{x^2\mathchar`-y^2}}(\bm{r})$ [Eq.~(\ref{eq:pair_cor})] for the bilayer Hubbard model.
\txcy{Figure}~\ref{fig:Pd_SPvsNP} shows the results of $P_{d_{x^2\mathchar`-y^2}}(\bm{r})$ with and without \txcy{the} spin quantum-number projection $\mathcal{L}^{S}$ for $L=20$.
The real space dependence of \txcy{these} supersonducting correlations shows quantitatively same behavior
and the long-distance averages of supersonducting correlations $\bar{P}_{d_{x^2\mathchar`-y^2}}$ are $0.048221\pm0.00001$ with $\mathcal{L}^{S}$ and $0.048856\pm0.00001$ without $\mathcal{L}^{S}$.
Therefore, \txcy{we conclude that} the spin quantum-number projection does not affect the superconducting \txcy{correlations}
for the bilayer Hubbard \txcy{hamiltonian}.}

\bibliography{iwano_bilayer}

\end{document}